\pdfoutput=1
 \documentclass[aps,prl,twocolumn,showpacs,superscriptaddress,preprintnumbers]{revtex4-1}  
\usepackage[normalem]{ulem}
\usepackage{amsmath}
\usepackage{amssymb}
\usepackage{epsfig}
\usepackage{graphicx}
\usepackage{hyperref}
\usepackage{tabu}
\usepackage{boldline}
\usepackage{xspace}
\usepackage{slashed}
\usepackage{multirow}
\usepackage{diagbox}
\usepackage{tabularx}
\usepackage{comment}

\usepackage{floatrow}
 
\newfloatcommand{capbtabbox}{table}[][\FBwidth]

\usepackage{color}
\usepackage[normal]{subfigure}
\usepackage[table]{xcolor}
\usepackage{enumitem}
\usepackage[utf8]{inputenc}
\usepackage{colortbl}
\definecolor{nicered}{rgb}{0.7,0.1,0.1}
\definecolor{nicegreen}{rgb}{0.1,0.5,0.1}
\definecolor{red}{rgb}{1.0, 0, 0}
\definecolor{pink}{RGB}{255, 0, 145}

\usepackage{lineno}

\definecolor{LightCyan}{rgb}{0.88,1,1}
\definecolor{piggypink}{rgb}{0.99, 0.87, 0.9}
\definecolor{applegreen}{rgb}{0.55, 0.71, 0.0}
\definecolor{darkpastelgreen}{rgb}{0.01, 0.75, 0.24}
\definecolor{green-yellow}{rgb}{0.68, 1.0, 0.18}

\newcommand{\beq}{\begin{equation}}
\newcommand{\eeq}{\end{equation}}
\newcommand{\beqa}{\begin{eqnarray}}
\newcommand{\eeqa}{\end{eqnarray}}

\newcommand{\unit}[1]{\ensuremath{\mathrm{\,#1}}\xspace}
\newcommand{\units}[1]{\unit{#1}}
\newcommand{\e}{\unit{e^{-}}}

    \newcommand{\Rs}{$R_{1e^-}$}
    \newcommand{\Rtwo}{$R_{2e^-}$}
    \newcommand{\Rthree}{$R_{3e^-}$}
    \newcommand{\Rfour}{$R_{4e^-}$}

\graphicspath{{figs/}}

\begin{document}


\title{SENSEI: Direct-Detection Results on sub-GeV Dark Matter \\ from a New Skipper-CCD
}

\author{The SENSEI Collaboration: \\ Liron Barak}

\affiliation{\normalsize\it 
 School of Physics and Astronomy, 
 Tel-Aviv University, Tel-Aviv 69978, Israel}

 \author{Itay M. Bloch}
 \affiliation{\normalsize\it 
 School of Physics and Astronomy, 
 Tel-Aviv University, Tel-Aviv 69978, Israel}

\author{Mariano Cababie}
\affiliation{\normalsize\it 
Department of Physics, FCEN, University of Buenos Aires and IFIBA, CONICET, Buenos Aires, Argentina}
\affiliation{\normalsize\it 
Fermi National Accelerator Laboratory, PO Box 500, Batavia IL, 60510, USA}

\author{Gustavo Cancelo}
\affiliation{\normalsize\it 
Fermi National Accelerator Laboratory, PO Box 500, Batavia IL, 60510, USA}

\author{Luke Chaplinsky}
\affiliation{\normalsize\it 
C.N.~Yang Institute for Theoretical Physics, Stony Brook University, Stony Brook, NY 11794, USA}
\affiliation{\normalsize\it 
Department of Physics and Astronomy, Stony Brook University, Stony Brook, NY 11794, USA}

\author{Fernando Chierchie}
\affiliation{\normalsize\it 
Fermi National Accelerator Laboratory, PO Box 500, Batavia IL, 60510, USA}

\author{Michael Crisler}
\affiliation{\normalsize\it 
Fermi National Accelerator Laboratory, PO Box 500, Batavia IL, 60510, USA}

\author{Alex Drlica-Wagner}
\affiliation{\normalsize\it 
Fermi National Accelerator Laboratory, PO Box 500, Batavia IL, 60510, USA}
\affiliation{\normalsize\it Kavli Institute for Cosmological Physics, University of Chicago, Chicago, IL 60637, USA}
\affiliation{\normalsize\it  Department of Astronomy and Astrophysics, University of Chicago, Chicago IL 60637, USA}

 \author{Rouven Essig}
\affiliation{\normalsize\it 
C.N.~Yang Institute for Theoretical Physics, Stony Brook University, Stony Brook, NY 11794, USA}

 \author{Juan Estrada}
\affiliation{\normalsize\it 
Fermi National Accelerator Laboratory, PO Box 500, Batavia IL, 60510, USA}

\author{Erez Etzion}
\affiliation{\normalsize\it 
 School of Physics and Astronomy, 
 Tel-Aviv University, Tel-Aviv 69978, Israel}

\author{Guillermo Fernandez Moroni}
\affiliation{\normalsize\it 
Fermi National Accelerator Laboratory, PO Box 500, Batavia IL, 60510, USA}

\author{Daniel Gift}
\affiliation{\normalsize\it 
C.N.~Yang Institute for Theoretical Physics, Stony Brook University, Stony Brook, NY 11794, USA}
\affiliation{\normalsize\it 
Department of Physics and Astronomy, Stony Brook University, Stony Brook, NY 11794, USA} 

\author{Sravan Munagavalasa}
\affiliation{\normalsize\it 
C.N.~Yang Institute for Theoretical Physics, Stony Brook University, Stony Brook, NY 11794, USA}
\affiliation{\normalsize\it 
Department of Physics and Astronomy, Stony Brook University, Stony Brook, NY 11794, USA}

 \author{Aviv Orly}
\affiliation{\normalsize\it 
 School of Physics and Astronomy, 
 Tel-Aviv University, Tel-Aviv 69978, Israel}

\author{Dario Rodrigues}
\affiliation{\normalsize\it 
Department of Physics, FCEN, University of Buenos Aires and IFIBA, CONICET, Buenos Aires, Argentina}
\affiliation{\normalsize\it 
Fermi National Accelerator Laboratory, PO Box 500, Batavia IL, 60510, USA}

\author{Aman Singal}
\affiliation{\normalsize\it 
Department of Physics and Astronomy, Stony Brook University, Stony Brook, NY 11794, USA}

 \author{Miguel Sofo Haro}
\affiliation{\normalsize\it 
Fermi National Accelerator Laboratory, PO Box 500, Batavia IL, 60510, USA}
\affiliation{Centro At\'omico Bariloche, CNEA/CONICET/IB, Bariloche, Argentina}

\author{Leandro Stefanazzi}
\affiliation{\normalsize\it 
Fermi National Accelerator Laboratory, PO Box 500, Batavia IL, 60510, USA}

\author{Javier Tiffenberg}
\affiliation{\normalsize\it 
Fermi National Accelerator Laboratory, PO Box 500, Batavia IL, 60510, USA}

\author{Sho Uemura}
\affiliation{\normalsize\it 
 School of Physics and Astronomy, 
 Tel-Aviv University, Tel-Aviv 69978, Israel}

\author{Tomer Volansky}
\affiliation{\normalsize\it 
 School of Physics and Astronomy,   
 Tel-Aviv University, Tel-Aviv 69978, Israel}

\author{Tien-Tien Yu}
\affiliation{\normalsize\it 
Department of Physics and Institute for Fundamental Science, University of Oregon, Eugene, Oregon 97403, USA}

\preprint{YITP-SB-2020-6, FERMILAB-PUB-20-158-AE-E}
\begin{abstract}

We present the first direct-detection search for sub-GeV dark matter using a new $\sim$2-gram high-resistivity Skipper-CCD from a dedicated fabrication batch that was optimized for dark-matter searches.  
Using 24 days of data acquired in the MINOS cavern at the Fermi National Accelerator Laboratory, we measure the lowest rates in silicon detectors of events containing one, two, three, or four electrons, and achieve world-leading sensitivity for a large range of sub-GeV dark matter masses. Data taken with different thicknesses of the detector shield suggest a correlation between the rate of high-energy tracks and the rate of single-electron events previously classified as ``dark current.'' 
We detail key characteristics of the new Skipper-CCDs, which augur well for the planned construction of the $\sim$100-gram SENSEI experiment at SNOLAB. 
\end{abstract}

\maketitle

\noindent\textbf{INTRODUCTION.}  
Dark matter (DM) candidates with masses below $\sim$1~GeV are well-motivated and have received increased attention in the past several years~\cite{Battaglieri:2017aum}.  However, such DM remains poorly constrained with direct-detection experiments, since 
the energy of the recoiling nucleus in searches for elastic DM-nucleus scattering is typically below detector thresholds. Instead, DM interactions with electrons can probe sub-GeV masses~\cite{Essig:2011nj}. 
The goal of the Sub-Electron-Noise Skipper-CCD Experimental Instrument (SENSEI) is to use ultralow-noise silicon Skipper-Charge-Coupled-Devices (Skipper-CCDs)~\cite{Tiffenberg:2017aac,skipper_2012} to probe DM down to masses of $\sim$500~keV scattering off electrons~\cite{Essig:2011nj,Essig:2015cda,Lee:2015qva,Graham:2012su} and DM down to masses of the silicon band gap of $\sim$1.2~eV being absorbed by electrons~\cite{An:2014twa,Bloch:2016sjj,Hochberg:2016sqx}. 
SENSEI can also probe DM-nucleus scattering through the `Migdal' effect~\cite{Ibe:2017yqa} down to $\sim$1~MeV~\cite{Essig:2019xkx}. 

An electron that is excited from the silicon valence band to the conduction band in one of the pixels of the Skipper-CCD 
typically creates one additional electron-hole pair (below, simply referred to as an ``electron'' and denoted as ``\e'') for each 3.8~eV of excitation energy above the band gap~\cite{Vavilov1962}.   DM-electron scattering typically creates only one to a few \e~\cite{Essig:2015cda}. The charge in each pixel is then moved pixel-to-pixel to a readout stage located in one of the corners of the Skipper-CCD, where the pixel charge is measured repeatedly and non-destructively to sub-electron-noise precision~\cite{Tiffenberg:2017aac}. 

The results presented in this paper are based on data collected with high-resistivity Skipper-CCDs procured in April 2019 
(our previous results used prototypes~\cite{Tiffenberg:2017aac,Crisler:2018gci,Abramoff:2019dfb}).  
The Skipper-CCD was designed at LBNL and fabricated at Teledyne DALSA Semiconductor.  
We describe key properties of these Skipper-CCDs, 
and describe data that suggest a correlation between the single-electron event rate, denoted as \Rs, and the environmental background rate as inferred from ``high-energy'' events, i.e., events with energies above 500~eV.  We also present new DM constraints using a blinded dataset collected from Feb.~25, 2020 to March~19, 2020 with a single Skipper-CCD placed $\sim$104~m~\cite{Adamson:2007gu} underground in the MINOS cavern at Fermi National Accelerator Laboratory (FNAL). Supplemental Materials (SM) contain additional details. 

\noindent\textbf{THE NEW SENSEI SKIPPER-CCD DETECTORS.}
The science-grade Skipper-CCDs consist of silicon with a resistivity of 18~k$\Omega$-cm, an active area of 9.216~cm $\times$ 1.329~cm, a thickness of $675~\mu\textrm{m}$, an active mass of 1.926~g, and 5443584~pix. 
No thinning process was applied to the back side to maximize the target mass and reduce fabrication cost.
Each Skipper-CCD has four identical amplifiers, one in each corner, which can read the entire CCD. However, the usual mode of operation is to read one quarter of the CCD consisting of 3072 rows and 443 columns of pixels.
The serial register for one quadrant, which is the first row of pixels that transfers the charge to the readout stage, is along the short side of the CCD and thus consists of 443 columns. 
When moving charge pixel-to-pixel in the serial register, random 1\e-events (``spurious charge'') are generated, which we measure to be $(1.664\pm 0.122)\times 10^{-4}$\e/pix and subtract from the observed \Rs\ (see SM). Each pixel has a volume of $15~\mu\textrm{m} \times 15~\mu\textrm{m} \times 675~\mu\textrm{m}$ and a mass of $3.537\times 10^{-7}$~g. 
The DM science data are taken with the output transistor of the amplifiers turned off during exposure, although we find no evidence for amplifier-induced events that occurred in the prototype detectors~\cite{Abramoff:2019dfb}, likely due to the improved quality of the silicon.  

\begin{figure}[t!]
\includegraphics[width=0.99\textwidth]{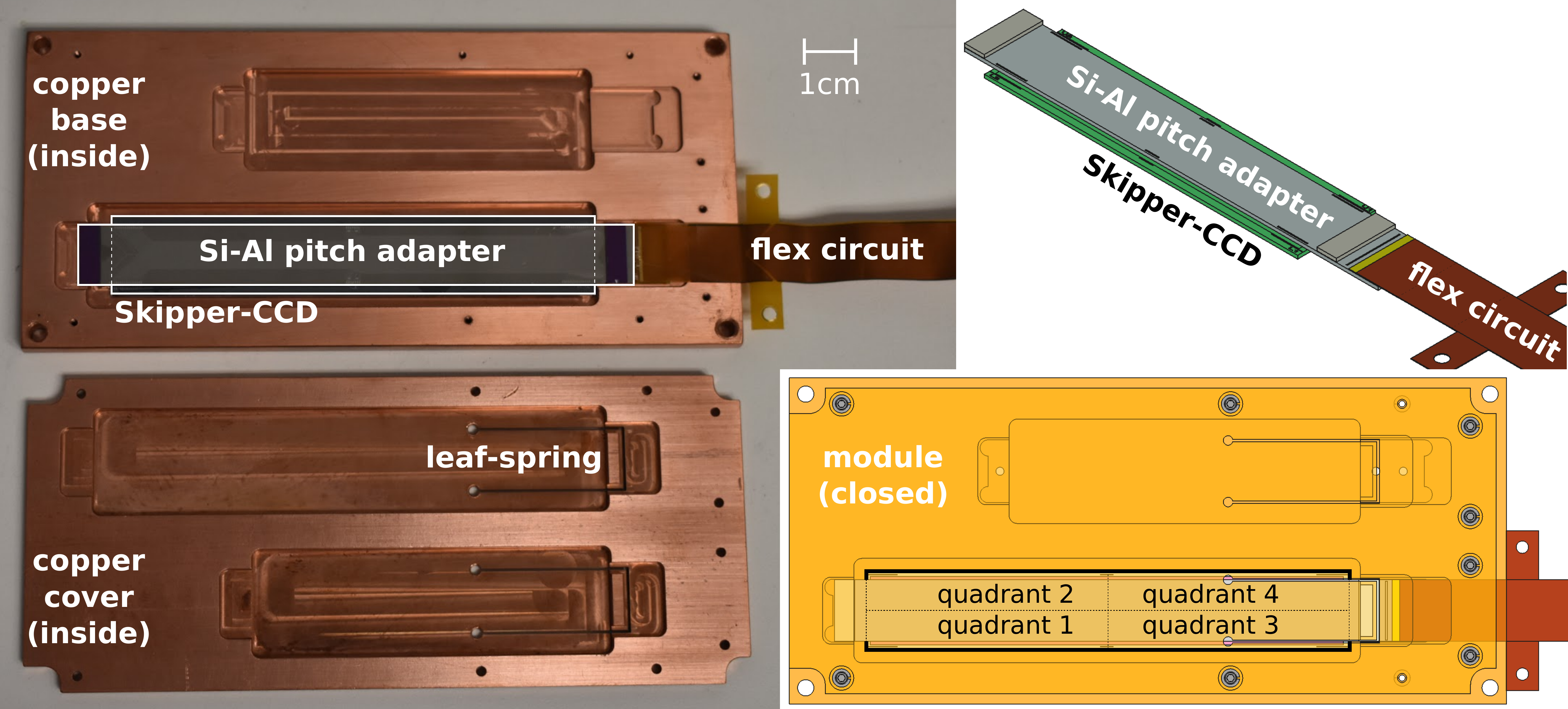}
\caption{A copper-Kapton flex circuit is laminated to a silicon-aluminum pitch adapter that is glued and wirebonded to the Skipper-CCD (top right); this is placed in a copper tray (top left), where a copper leaf-spring (bottom left) maintains constant pressure for consistent thermal contact between the CCD and the tray when closed inside the module (shown transparent, bottom right). 
\vspace{-4mm}
}
\label{fig:module}
\end{figure}

A silicon-aluminum pitch adapter and copper-Kapton flex cable were glued and wirebonded to the CCD. The overall width of this assembly is 
no larger than the width of the CCD, allowing dense packing for large-scale Skipper-CCD experiments. This assembly was placed in a copper tray, where a copper leaf-spring maintains constant pressure for consistent thermal contact between the CCD and the tray (Fig.~\ref{fig:module}).
The module was placed in the same vessel used for the results in~\cite{Abramoff:2019dfb}, but with extra lead shielding placed around the vessel (see SM), which reduced the high-energy event rate and \Rs\ (see below). The CCD was operated at a temperature of 135~K.

The readout and control systems are fully integrated in a new single-board electronics optimized for Skipper-CCD sensors. This Low-Threshold-Acquisition system (LTA)~\cite{LTA8709274,Cancelo:2020egx} provides a flexible and scalable solution for detectors with target masses up to a few hundred grams. The root-mean-square single-sample readout noise is 2.5\e.

\begin{figure}[t!]
\begin{center}
\includegraphics[width=0.99\textwidth]{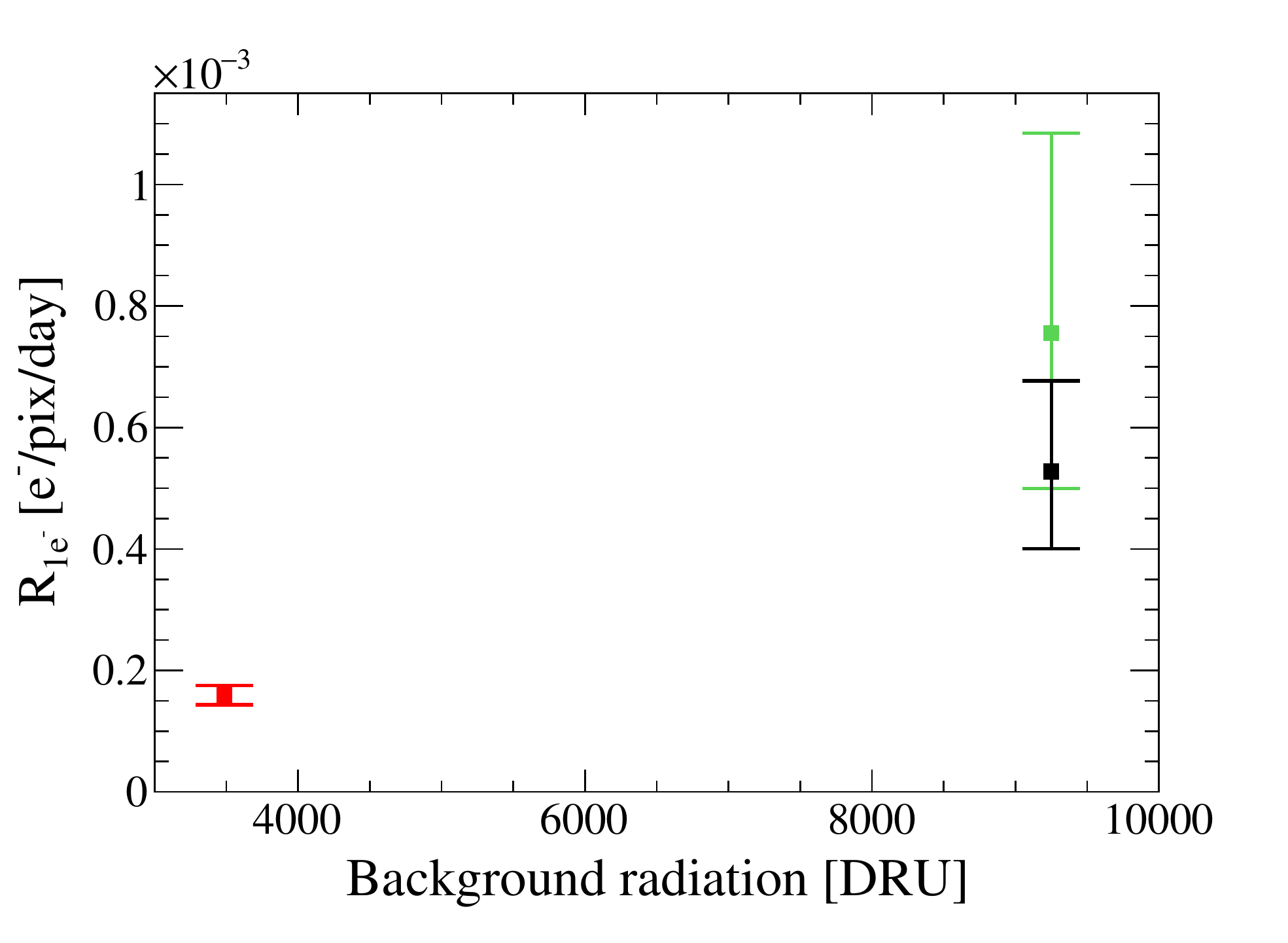}
\end{center}
\caption{Single-electron event rate, \Rs\ (after subtracting the spurious charge) versus the rate of events with energies 500~eV to 10~keV with extra shielding (red) and without extra shielding for one image with the amplifier off during exposure (green) and when combined with three images with the amplifier on during exposure (black). 
\vspace{-4mm}
}
\label{fig:Background}
\end{figure}

\noindent\textbf{DEPENDENCE OF SINGLE-ELECTRON RATE ON ENVIRONMENTAL BACKGROUNDS.}
We find evidence for a correlation between 
the rate of high-energy background events and \Rs.  In the DM science data (see below), which has extra lead shielding, the rate of events with 500~eV to 10~keV energy is 3370 DRU (1~DRU is 1~event/kg/day/keV), while $R_{1e^-}=(1.594 \pm 0.160)\times 10^{-4}$~\e/pix/day, i.e.~$(450\pm 45)$/g-day. This is the smallest \Rs\ achieved with a semiconductor target. In one image taken without the extra lead (the ``standard'' shield), we find $R_{1e^-}=(7.555^{+3.286}_{-2.562})\times 10^{-4}$~\e/pix/day. Three additional standard-shield images (but taken with the amplifier voltages turned on during exposure) show $R_{1e^-}=(4.302^{+1.743}_{-1.426})\times 10^{-4}$~\e/pix/day, so that, averaged over the four images,  $R_{1e^-}=(5.312^{+1.490}_{-1.277})\times 10^{-4}$~\e/pix/day, i.e.~$1492^{+421}_{-361}$/g-day.  The combined standard-shield high-energy background rate is 9700~DRU; see Fig.~\ref{fig:Background} and SM for more details. 
The origin of the 1\e-events requires further study. 
We have insufficient data to measure \Rtwo\ for the standard-shield case. The high-energy event spectra are shown in the SM.

\noindent\textbf{DATA COLLECTION.}
We collect blinded data for constraining DM that produces events with $\le 4$ electrons. 
We expose the Skipper-CCD for 20 hours, and then read each quadrant through one amplifier with 300 samples per pixel. We refer to one such exposure-and-readout as an ``image.''  We took 22 images of DM science data before a mandatory shutdown. All charge on the CCD is erased before taking a new image.  The read time per sample is 42.825~$\mu$s, while the readout time of the entire active area is 5.153~hours. 
Commissioning data, consisting of (7) 20-hour-exposure images, were used to determine the data quality cuts.  
During commissioning, ``quadrant-1'' and ``quadrant-2'' showed excellent performance, with a root-mean square noise of 0.146\e and 0.139\e (with negligible error bars), respectively. 
``Quadrant-4'' had an excessively high charge-transfer-inefficiency (consistent with a disconnected serial register clock) and its data were discarded. In addition, ``quadrant-3'' (next to quadrant-4 on the short CCD-side), with a noise of 0.142\e, had an excess of 1\e events in the entire quadrant, but especially in the first $\sim$100 columns, consistent with possible blackbody radiation from the surrounding warm vessel leaking onto that part of the cold CCD through the leaf-spring slots (Fig~\ref{fig:module}).  Before unblinding, we thus discarded quadrant-3's data for the 1\e and 2\e analyses; however, we include its columns 93 to 443 for the 3\e and 4\e analyses to increase our exposure and since the expected probability of a single spurious 3\e event is at the percent level. The total exposure (before cuts) of the DM search data is 19.93~g-day for the 1\e and 2\e analyses, and 27.82~g-day for the 3\e and 4\e analyses.

\begin{table}[t!]
\begin{center}
\begin{footnotesize}
\begin{tabular}{|l||*{8}{c|}}
\hline
\diagbox{Cuts}{ $N_{e}$} & \multicolumn{2}{c|}{1} & \multicolumn{2}{c|}{2} & \multicolumn{2}{c|}{3}        & \multicolumn{2}{c|}{4}             \\ \hline \hline
1.~Charge Diffusion            & \multicolumn{2}{c|}{1.0}   & \multicolumn{2}{c|}{$0.228$} & \multicolumn{2}{c|}{$0.761$}     & \multicolumn{2}{c|}{$0.778$}       \\ \hline
\hline
         & Eff.   & \#Ev & Eff.   & \#Ev & Eff.   & \#Ev & Eff.   & \#Ev      \\ \hline
2.~Readout Noise         & 1   & $>10^5$ & 1   & 58547 & 1 & 327 & 1 & 155      \\ 
3.~Crosstalk                & 0.99 & $>10^5$ & 0.99 & 58004 & 0.99 & 314 & 0.99 & 153    \\ 
4.~Serial Register      & $\sim 1$ & $>10^5$ & $\sim 1$ & 57250 & $\sim 1$ & 201 & $\sim 1$ & 81    \\ 
5.~Low-E Cluster       & 0.94 & 42284 & 0.94 & 301 & 0.69 & 35 & 0.69 & 7    \\ 
6.~Edge                     & 0.70 & 25585 & 0.90 & 70 & 0.93 & 8 & 0.93 & 2    \\ 
7.~Bleeding Zone            & 0.60 & 11317 & 0.79 & 36 & 0.87 & 7 & 0.87 & 2    \\ 
8.~Bad Pixel/Col.         & 0.98 & 10711 & 0.98 & 24 & 0.98 & 2 & 0.98 & 0    \\ 
9.~Halo                     & 0.18 & 1335 & 0.81 & 11 & $\sim 1$ & 2 & $\sim 1$ & 0      \\ 
10.~Loose Cluster          & \multicolumn{2}{c|}{N/A}    & 0.89 & 5 & 0.84 & 0 & 0.84 & 0    \\
11.~Neighbor        & $\sim 1$ & 1329 & $\sim 1$ & 5 & \multicolumn{4}{c|}{N/A}      \\ \hline  
Total Efficiency            & \multicolumn{2}{c|}{0.069} & \multicolumn{2}{c|}{0.105} & \multicolumn{2}{c|}{0.341}      & \multicolumn{2}{c|}{0.349}      \\ \hline
Eff.~Efficiency            & \multicolumn{2}{c|}{0.069} & \multicolumn{2}{c|}{0.105} & \multicolumn{2}{c|}{0.325}      & \multicolumn{2}{c|}{0.327}      \\ \hline
Eff.~Exp.~[g-day]          & \multicolumn{2}{c|}{1.38} & \multicolumn{2}{c|}{2.09} & \multicolumn{2}{c|}{9.03}      & \multicolumn{2}{c|}{9.10}      \\ \hline \hline
Observed Events            & \multicolumn{2}{c|}{1311.7$^{(*)}$}      & \multicolumn{2}{c|}{5}     & \multicolumn{2}{c|}{0}      & \multicolumn{2}{c|}{0}      \\ \hline
90\%CL [g-day]$^{-1}$ & \multicolumn{2}{c|}{525.2$^{(*)}$} & \multicolumn{2}{c|}{4.449} 
                                    & \multicolumn{2}{c|}{0.255} & \multicolumn{2}{c|}{0.253}      \\ \hline
\end{tabular}
\end{footnotesize}
\caption{Efficiencies and number of events containing 1\e, 2\e, 3\e, or 4\e events that pass the masking cuts for the 1\e, 2\e, 3\e, and 4\e analysis, respectively. 
The Charge Diffusion cut assumes the DM generates single-pixel events for $N_e=2\e$ or contiguous multi-pixel events for $N_e=3\e$ or $N_e=4\e$. 
The Total Efficiency is the fraction of pixels that pass all cuts, while the Effective (``Eff.'') Efficiency is exposure-corrected (since each pixel has a unique exposure) and, for 3\e and 4\e, includes a geometric efficiency.  The bottom three rows respectively list the efficiency-corrected exposure, the number of observed events after cuts, and the 90\% CL~limits.  $^{(*)}$For 1\e, we list the number after subtracting 0\e (adding 1\e) events that are mis-classified as 1\e (0\e or 2\e); the quoted limit is after subtracting the spurious charge. 
} 
\label{tab:eff}
\end{center}
\end{table}%
\noindent\textbf{DATA ANALYSIS.}
We perform four analyses: on 1\e events, on single-pixel 2\e events, and on events consisting of a \textit{contiguous} set of pixels containing a total of 3\e or 4\e. Most event selection criteria are common to the four analyses, but there are important differences, mostly because the 1\e and (to a lesser extent) the 2\e analyses are not exposure-limited. Due to nonzero noise, we define a (1\e, 2\e, 3\e, 4\e) pixel to have a \textit{measured} charge in the range ((0.63,1.63], (1.63,2.5], (2.5,3.5], (3.5,4.5])\e, respectively. Pixels with $\ge$1\e have a measured charge of $>$0.63\e.  A ``cluster'' is defined as a contiguous set of neighboring pixels that each have a measured charge of $>$0.63\e. Given a pixel, a ``neighboring'' pixel is one of the eight adjacent pixels. 
The charge of a cluster is the sum of the pixel charges.  For counting the final number of 1\e events, we use a Gaussian fit to remove 0\e (add 1\e) events that have a measured charge $>$0.63\e ($\le$0.63\e or $>$2.5\e). 

\begin{figure}[!t]
\begin{center}
 \includegraphics[width=0.99\textwidth]{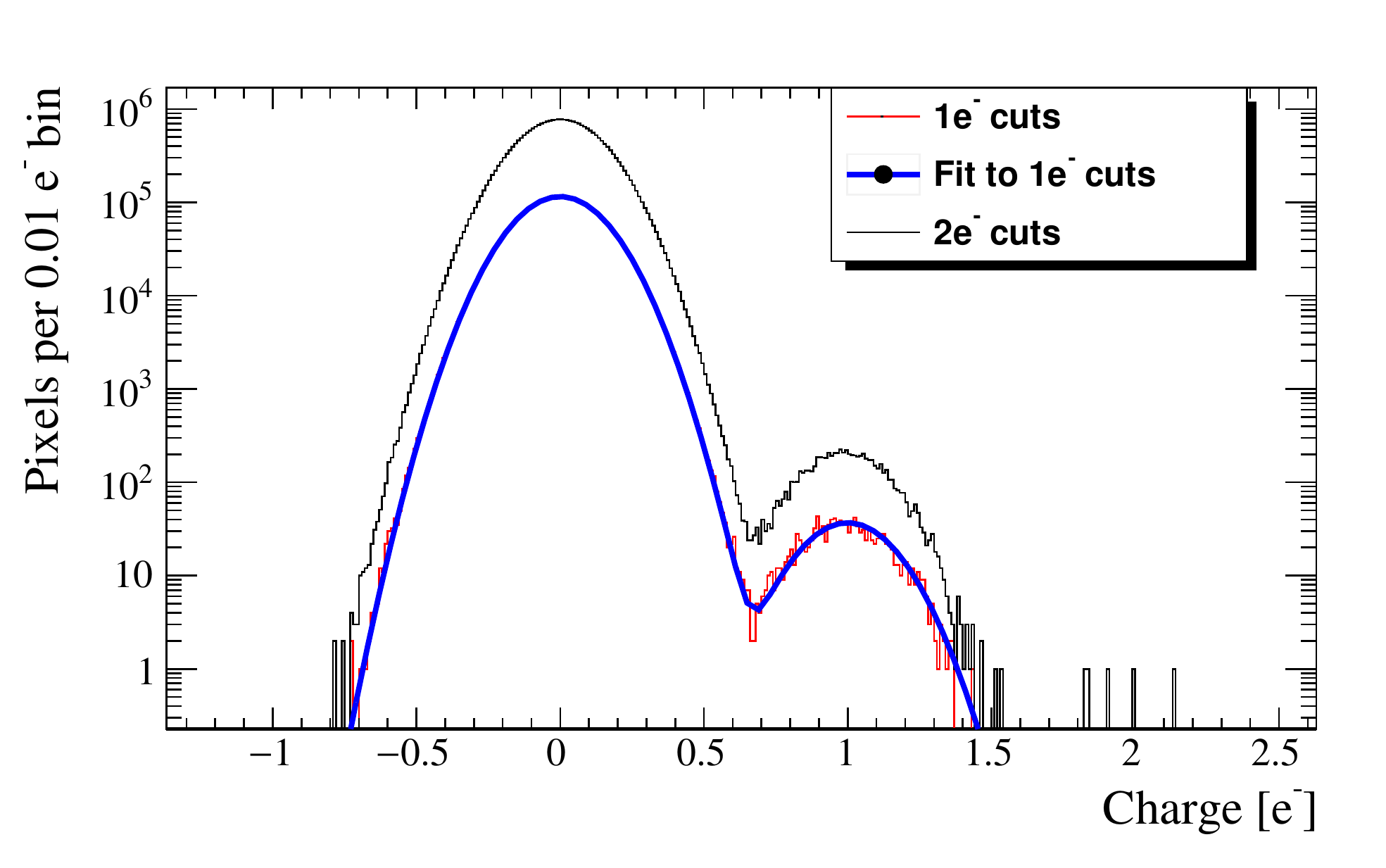}
 \end{center}
\caption{The pixel charge spectra (after selection cuts) used for the 1\e and 2\e analyses. A double-Gaussian fit is shown for the spectrum with 1\e cuts. There are no 3\e or 4\e events. 
}
\label{fig:spectra}
\end{figure}

\begin{nolinenumbers}
\begin{figure*}
\hskip -1cm 
\floatbox[{\capbeside\thisfloatsetup{capbesideposition={right,center},capbesidewidth=0.31\textwidth}}]{figure}[\FBwidth]
{\caption{
90\%~CL~constraints (cyan solid line) on: DM-\e cross section, $\overline{\sigma}_e$, 
versus DM mass, $m_\chi$, for two DM form factors, $F_{\rm DM}(q)=1$ (\textbf{top left}) and $F_{\rm DM}(q)=(\alpha m_e/q)^2$ (\textbf{top right});  DM-nucleus cross section, $\overline{\sigma}_n$, for a light mediator (\textbf{bottom left}); and the kinetic-mixing parameter, $\epsilon$, versus the dark-photon mass, $m_{A'}$, for dark-photon-DM absorption (\textbf{bottom right}). 
Constraints are shown on DM-\e scattering also from the SENSEI prototype~\cite{Crisler:2018gci,Abramoff:2019dfb}, XENON10/100~\cite{Essig:2017kqs}, DarkSide-50~\cite{Agnes:2018oej}, EDELWEISS~\cite{Arnaud:2020svb}, CDMS-HVeV~\cite{Agnese:2018col},
XENON1T~\cite{Aprile:2019xxb},
DAMIC~\cite{Aguilar-Arevalo:2019wdi}, solar reflection (assuming DM couples only to \e)~\cite{An:2017ojc}; constraints on DM-nucleus scattering from SENSEI, XENON10/100/1T~\cite{Essig:2019xkx} and LUX~\cite{Akerib:2018hck}; and constraints on absorption from SENSEI~\cite{Crisler:2018gci,Abramoff:2019dfb}, DAMIC~\cite{Aguilar-Arevalo:2016zop,Aguilar-Arevalo:2019wdi}, EDELWEISS~\cite{Arnaud:2020svb}, XENON10/100, CDMSlite~\cite{Bloch:2016sjj}, and the Sun~\cite{An:2013yfc,Redondo:2013lna,Bloch:2016sjj}. Orange regions are combined benchmark model regions for heavy~\cite{Boehm:2003hm, Essig:2011nj,Lin:2011gj,Essig:2015cda,Hochberg:2014dra,Kuflik:2017iqs,DAgnolo:2019zkf} and light~\cite{Essig:2011nj,Essig:2015cda,Chu:2011be,Dvorkin:2019zdi} mediators.
 \label{fig:DMresults}}}
{
\begin{minipage}{0.67\textwidth}
\begin{center}
\includegraphics[width=0.49\textwidth]{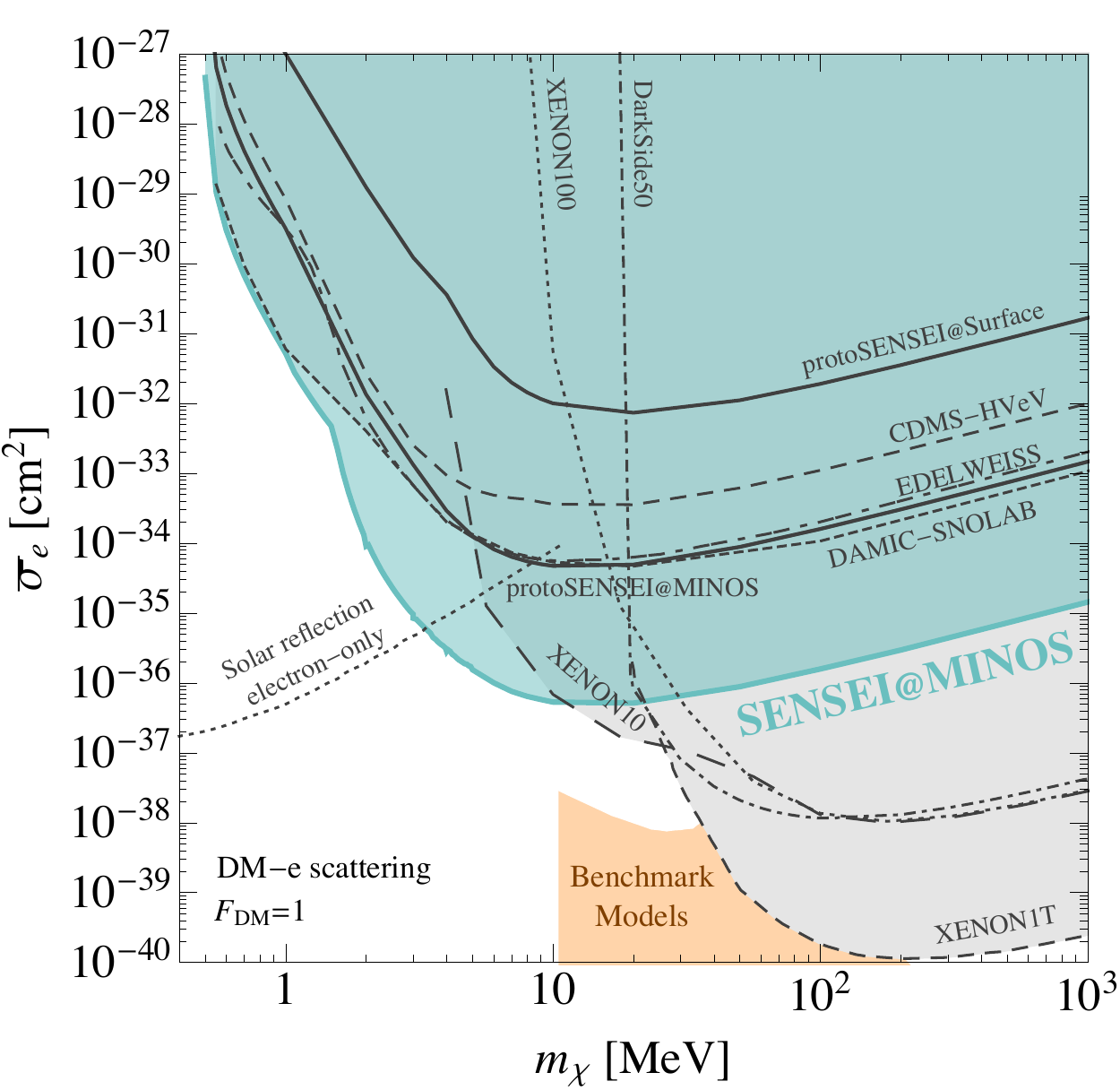}
 \includegraphics[width=0.49\textwidth]{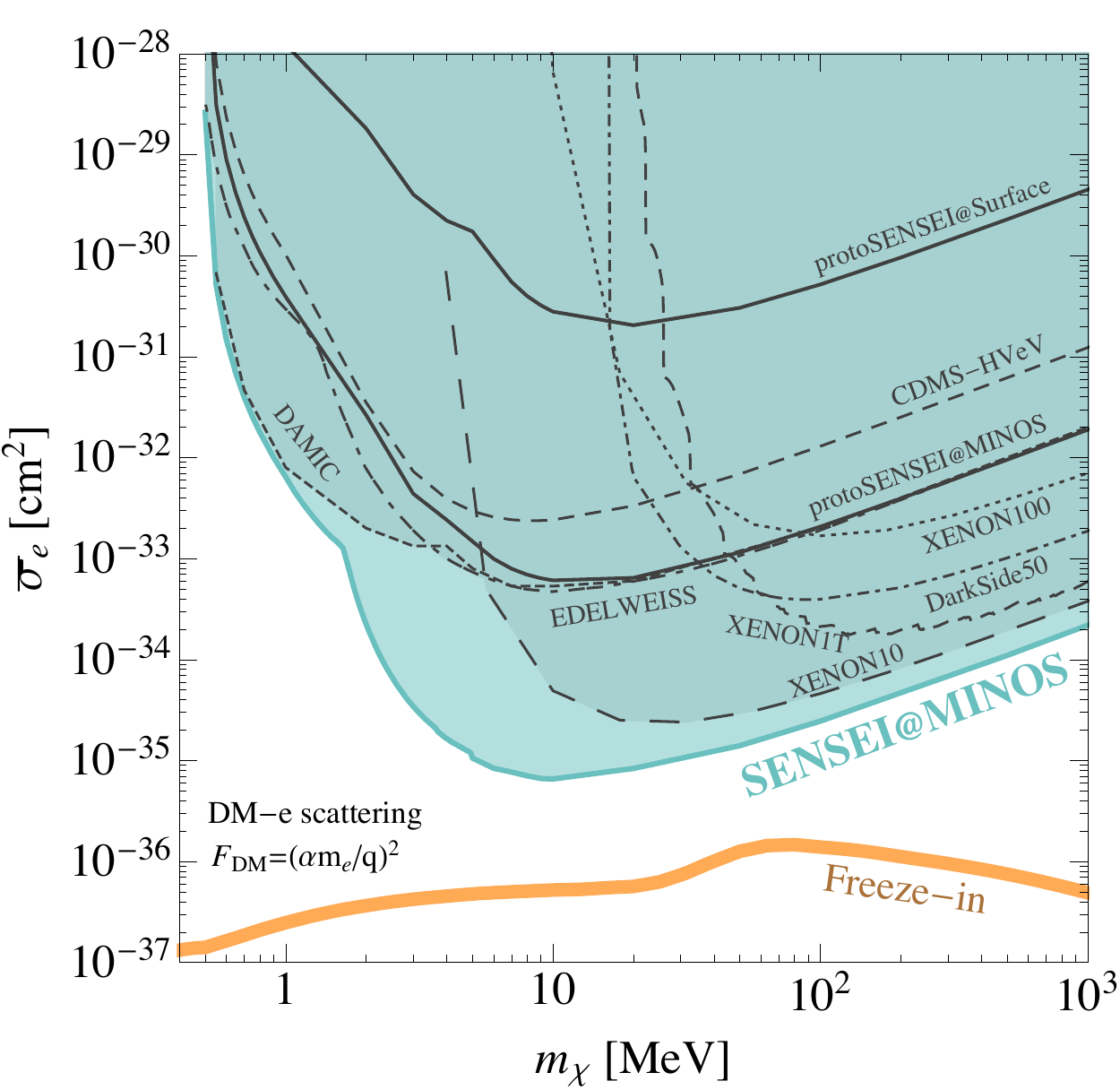}\\
\includegraphics[width=0.49\textwidth]{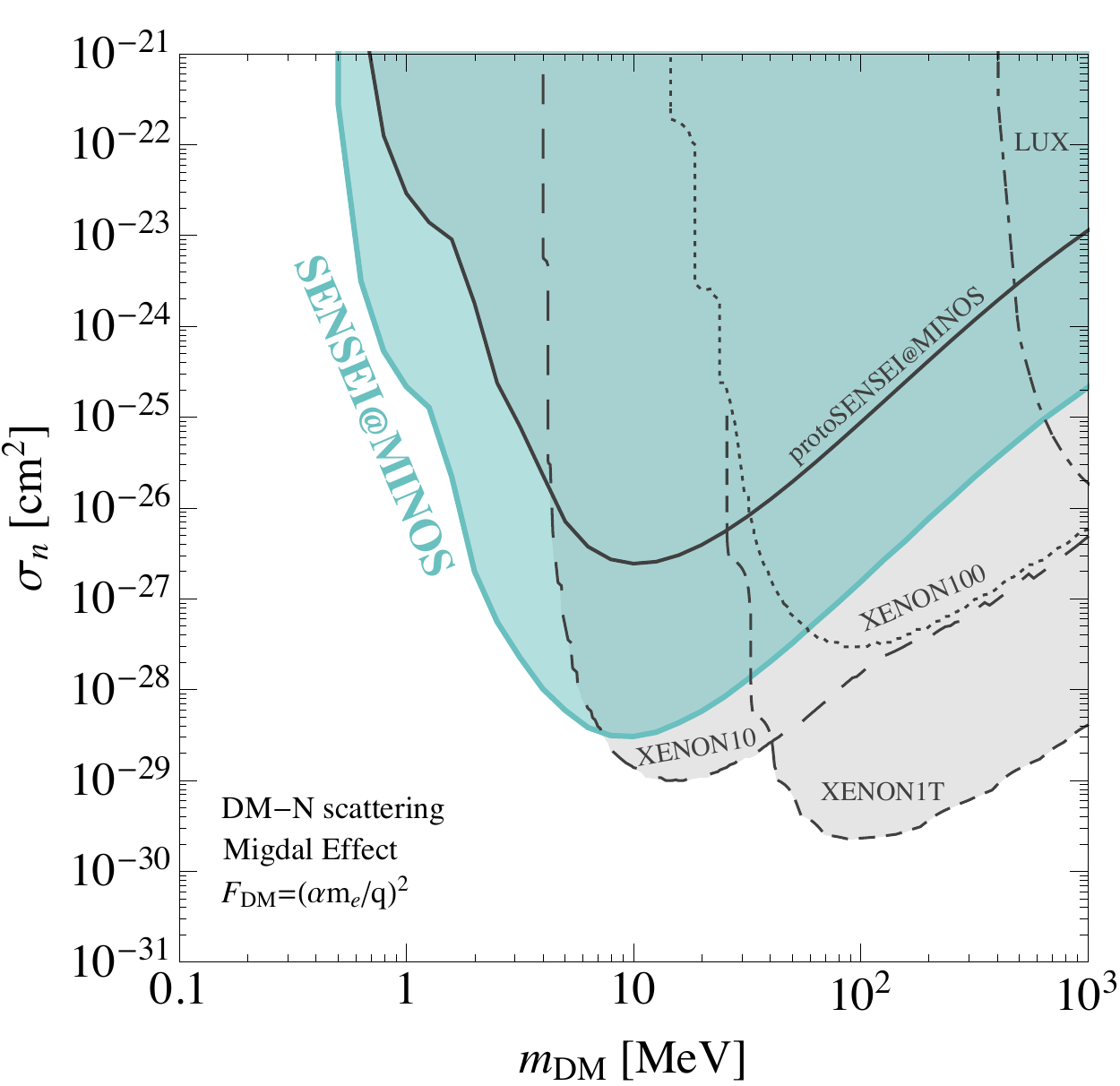}
\includegraphics[width=0.49\textwidth]{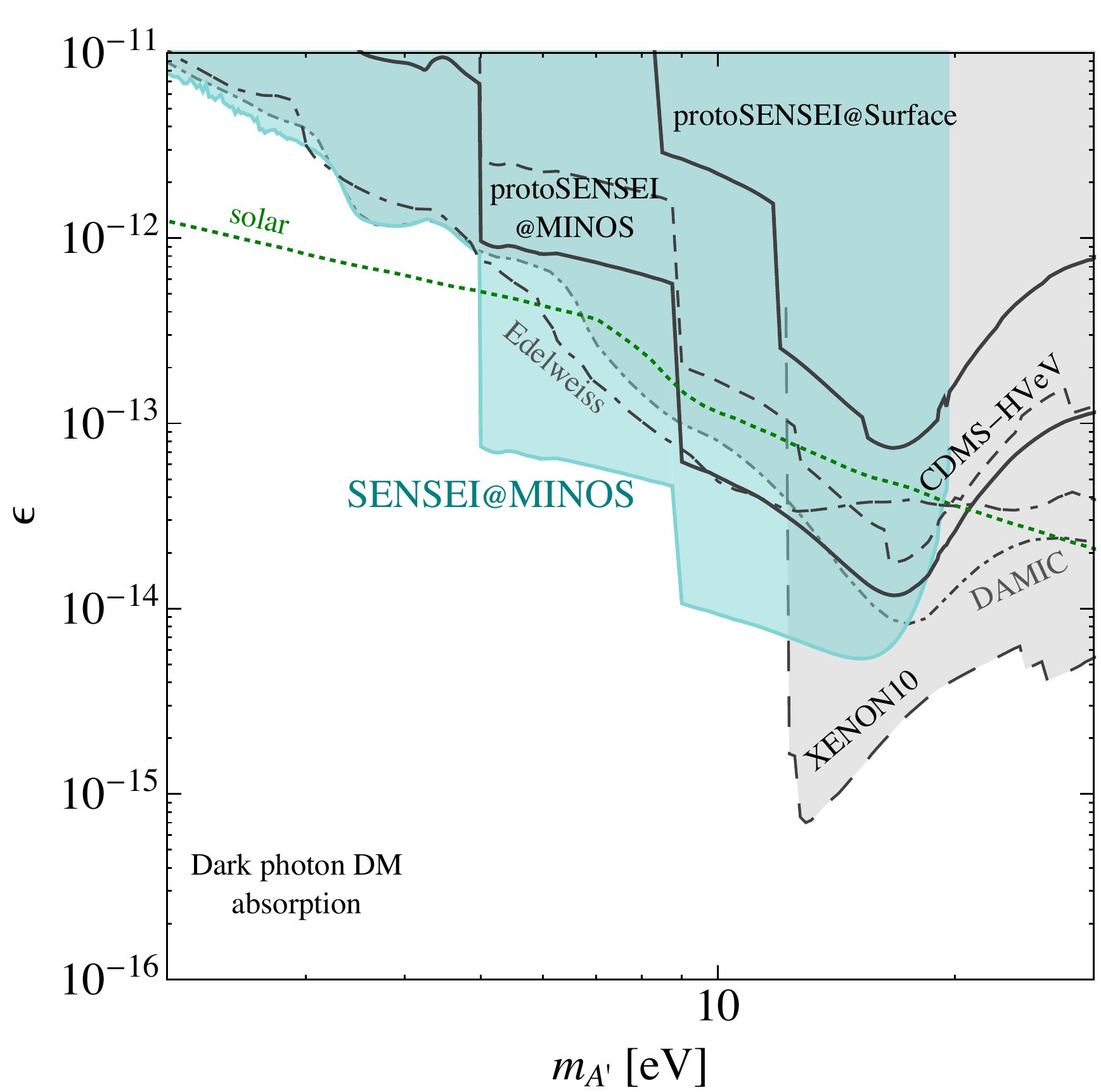}
\end{center}
\end{minipage}
}
\end{figure*}
\end{nolinenumbers}

We summarize next the event-selection criteria (see Table~\ref{tab:eff}). Additional details are in the SM.  
\begin{itemize}[leftmargin=*]\addtolength{\itemsep}{-0.7\baselineskip}
\item {\bf Charge Diffusion.} We account for the efficiency for DM events with $\ge$2\e to be spread out over more than one pixel due to charge diffusion.  The efficiency for detecting 1\e events is unity~\cite{LBNLQE}. 
\item {\bf Readout Noise.}  We veto images in which the readout noise is 30\% larger than the expected readout noise.  No such images are observed. 
\item {\bf Crosstalk.} 
A pixel is masked if it is read at the same time as another pixel containing $>$700~electrons. 
\item {\bf Serial Register Hit.} We remove isolated horizontal lines of charge, which indicate a background event that hit the serial register during readout. 
\item {\bf Low-Energy Cluster.} 
We observe disjoint groupings of $\ge$2\e events that are too close to be a coincidence.
For the 1\e and 2\e (3\e and 4\e) analyses, we thus remove four (20) pixels in all directions from all pixels that are part of a cluster containing at least 5\e (2\e). We do not remove the pixels of the cluster itself.
\item {\bf Edge Mask.}  We remove 60 (20) pixels around all edges of a quadrant for the 1\e ($\ge$2\e) analyses, which corresponds to the Halo Mask (described below) for any possible high-energy events occurring just outside of the quadrant. %
\item {\bf Bleeding Zone Mask.}
To avoid spurious events from charge-transfer inefficiencies, we mask 100 (50) pixels upstream in the vertical and horizontal direction of any pixel containing more than 100\e for the 1\e ($\ge$2\e) analyses. This distance is doubled for columns where we observe a high bleeding rate.
\item {\bf Bad Pixels and Bad Columns.} 
We further limit the impact of defects that cause charge leakage or charge-transfer inefficiencies by identifying and masking pixels and columns that have a significant excess of charge.
\item {\bf Halo Mask.}  Pixels with more than 100\e, from high-energy background events, correlate with an increased rate of low-energy events in nearby pixels.  We observe a monotonic decrease in \Rs\ as a function of the radial distance, $R$, from pixels with a large charge.  We mask pixels out to $R=60$~pix ($R=20$~pix) for the 1\e ($\ge$2\e) analyses. 
\item {\bf Loose Cluster Mask.}  
We find a correlation between the number of 1\e and 2\e events in regions of size $\sim$10$^3$~pix$^2$. Since there is no reason for a 2\e DM event to be spatially correlated with an excess of 1\e events, we mask regions with an excess of 1\e events. We apply this mask only for the $\ge$2\e analyses.
\item{\bf Neighbor Mask.} For the 1\e and 2\e DM analyses only, we require the DM signal to be contained in a single pixel and only select pixels whose eight neighboring pixels are empty. We thus mask all pixels that have a neighboring pixel with $\ge1\e$.
\end{itemize}%

The efficiencies of, and number of events passing, these selection cuts are given in Table~\ref{tab:eff}, which also shows the number of observed events and the inferred 90\% confidence-level (CL) upper limits on the rates. We assume that a DM signal is uniformly distributed across the CCD, so that a cut's efficiency on a DM signal is proportional to the loss in exposure from that cut.

\noindent\textbf{DARK MATTER RESULTS.}
The results for the four analyses are: 
\begin{itemize}[leftmargin=*]\addtolength{\itemsep}{-0.7\baselineskip}
\item \textbf{1e$^{-}$:} 
From the observed \Rs\ of $(3.363\pm 0.094) \times 10^{-4}$~\e/pix/day, we subtract the (exposure independent) spurious charge contribution of $(1.664 \pm 0.122)\times 10^{-4}$\e/pix, to arrive at a \Rs\ of 
$(1.594 \pm 0.160)\times 10^{-4}$\e/pix/day, or $(450\pm 45)$~events/g-day, where the errors have been added in quadrature.
For calculating a DM limit below, we conservatively take the 1311.7 observed 1\e events and subtract the $2\sigma$ lower limit on the number of expected spurious-charge events ($649-2\times 47.5=554$~events), arriving at $\sim$758~1\e-events.  The known contributions to \Rs\ that we do not subtract are environmental backgrounds and dark current (from thermal excitations). Of these, we expect the dark current contribution to be more than an order of magnitude lower than the observed \Rs\ (see SM).
\item \textbf{2e$^{-}$:} The 5 observed single-pixel 2\e events imply 
\Rtwo$=$2.399~events/g-day, with a 90\% CL~upper limit of \Rtwo$=$4.449/g-day ($\simeq 0.051$~Hz/kg).  This is more than two orders of magnitude lower than previous measurements of \Rtwo\ in solid-state detectors, and strongly disfavors a possible DM interpretation for the excess events observed in previous experiments~\cite{Kurinsky:2020dpb}. 
\item \textbf{3e$^{-}$ and 4e$^{-}$:} We observe zero 3\e and 4\e-clusters to find 90\% CL-upper limits on \Rthree of $0.255$/g-day and on \Rfour of 0.253/g-day. 
\end{itemize}

Fig.~\ref{fig:spectra} shows the observed spectra of events after all cuts. We use these data to constrain DM that scatters off electrons~\cite{Essig:2011nj,Essig:2015cda}, DM that is absorbed by electrons~\cite{An:2014twa,Bloch:2016sjj,Hochberg:2016sqx,HenkeExp,HenkeDatabase,EDWARDS1985547}), and DM that scatters off nuclei through the Migdal effect~\cite{Essig:2019xkx}. 
Current estimates of the Migdal effect at low recoil energies and especially for DM masses $\lesssim$10~MeV are uncertain~\cite{Essig:2019xkx}, so the resulting limits on DM-nucleus scattering should be viewed as approximate only.  
We assume an electron with recoil energy $E_e$ generates $(1+{\rm Floor}[(E_e-{1.2~{\rm eV}})/\varepsilon)$\e; new measurements at $\sim$6~keV find $\varepsilon=3.75~{\rm eV}$~\cite{Rodrigues2020}, but we will take $\varepsilon=3.8~{\rm eV}$ for consistency with other DM results in the literature and since the precise extrapolation to low energies remains uncertain. We conservatively ignore Fano-factor fluctuations for scattering, while for absorption we follow~\cite{Bloch:2016sjj}. We assume a local DM density of $\rho_{\rm DM}=0.3\units{GeV/cm^3}$~\cite{Bovy:2012tw}, a standard isothermal Maxwellian velocity distribution~\cite{Lewin:1995rx} with a DM escape velocity of 600~km/s, a mean local DM velocity of 230~km/s, and an average Earth velocity of 240~km/s.  
Fig.~\ref{fig:DMresults} shows the resulting ``SENSEI@MINOS'' 90\% c.l.~limit that combines the four analyses. We use a likelihood-ratio test based on~\cite{Cowan:2010js}, with a toy MC (instead of the asymptotic approximation) to compute the distribution of the $q_{\mu}$ statistics used for the calculation of the $p$-value.

For DM-electron scattering via a heavy (light) mediator, SENSEI@MINOS provides world-leading constraints for $m_\chi\sim 500$~keV--10~MeV ($m_\chi \gtrsim 500$~keV). For DM-nucleus scattering through a light mediator and for DM absorption on electrons, SENSEI@MINOS provides world-leading constraints for $m_\chi\sim 600$~keV--5~MeV and $m_\chi\sim 1.2$~eV--12.8~eV, respectively. 

\noindent\textbf{OUTLOOK.}
The SENSEI Collaboration is in the process of packaging and testing $\sim 75$ sensors from the same batch as the one used for this work. Up to $\sim$50 ($\sim$100~g) of packaged science-grade Skipper-CCDs will be deployed in a phased approach inside a low radiation shield currently being built at SNOLAB. 
We expect that the low radiation environment will translate to even lower \Rs. We plan to commission the first batch of sensors (tens of grams) over the next year. 
We plan to accumulate $\sim$100~g-years of exposure over 1 to 2~years.

Supplemental Materials provide additional details of our analysis, and contain additional references~\cite{janesick2001scientific,Haro:2016,Spurious-charge,Holland:2003,Aguilar-Arevalo:2016ndq, CONNIE_2019,Haro_2020,Bernstein:2017gsy,69907,Holland:2003,KRISHNAMOORTHY200423}.

\begin{acknowledgments}
\noindent\textbf{ACKNOWLEDGMENTS.}
We thank Kyle Cranmer for useful discussions on the limit calculation.  We are grateful for the support of the Heising-Simons Foundation under Grant No.~79921.
This work was supported by Fermilab under U.S.~Department of Energy (DOE) Contract No.~DE-AC02-07CH11359. 
The CCD development work was supported in part by the Director, Office of Science, of the DOE under No.~DE-AC02-05CH11231. RE acknowledges support from DOE Grant DE-SC0017938 and Simons Investigator in Physics Award~623940. 
The work of TV and EE is supported by the I-CORE Program of the Planning Budgeting Committee and the Israel Science Foundation (grant No.1937/12). TV is further supported  by the European Research Council (ERC) under the EU Horizon 2020 Programme (ERC- CoG-2015 -Proposal n.~682676 LDMThExp), and a grant from The Ambrose Monell Foundation, given by the Institute for Advanced Study.
The work of SU is supported in part by the Zuckerman STEM Leadership Program.
IB is grateful for the support of the Alexander Zaks Scholarship, The Buchmann Scholarship, and the Azrieli Foundation.
This manuscript has been authored by Fermi Research Alliance, LLC under Contract No. DE-AC02-07CH11359 with the U.S.~Department of Energy, Office of Science, Office of High Energy Physics. The United States Government retains and the publisher, by accepting the article for publication, acknowledges that the United States Government retains a non-exclusive, paid-up, irrevocable, world-wide license to publish or reproduce the published form of this manuscript, or allow others to do so, for United States Government purposes.
\end{acknowledgments}

\bibliographystyle{apsrev4-1}
\bibliography{main.arxiv.v3.bbl}
 
\clearpage

\begin{center}
    \begin{large} \textbf{SUPPLEMENTAL MATERIALS for:\\
``SENSEI: Direct-Detection Results on sub-GeV Dark Matter \\ from a New Skipper-CCD''}
    \end{large}
\end{center}

\begin{figure}[b!]
\begin{center}
\includegraphics[width=0.99\textwidth]{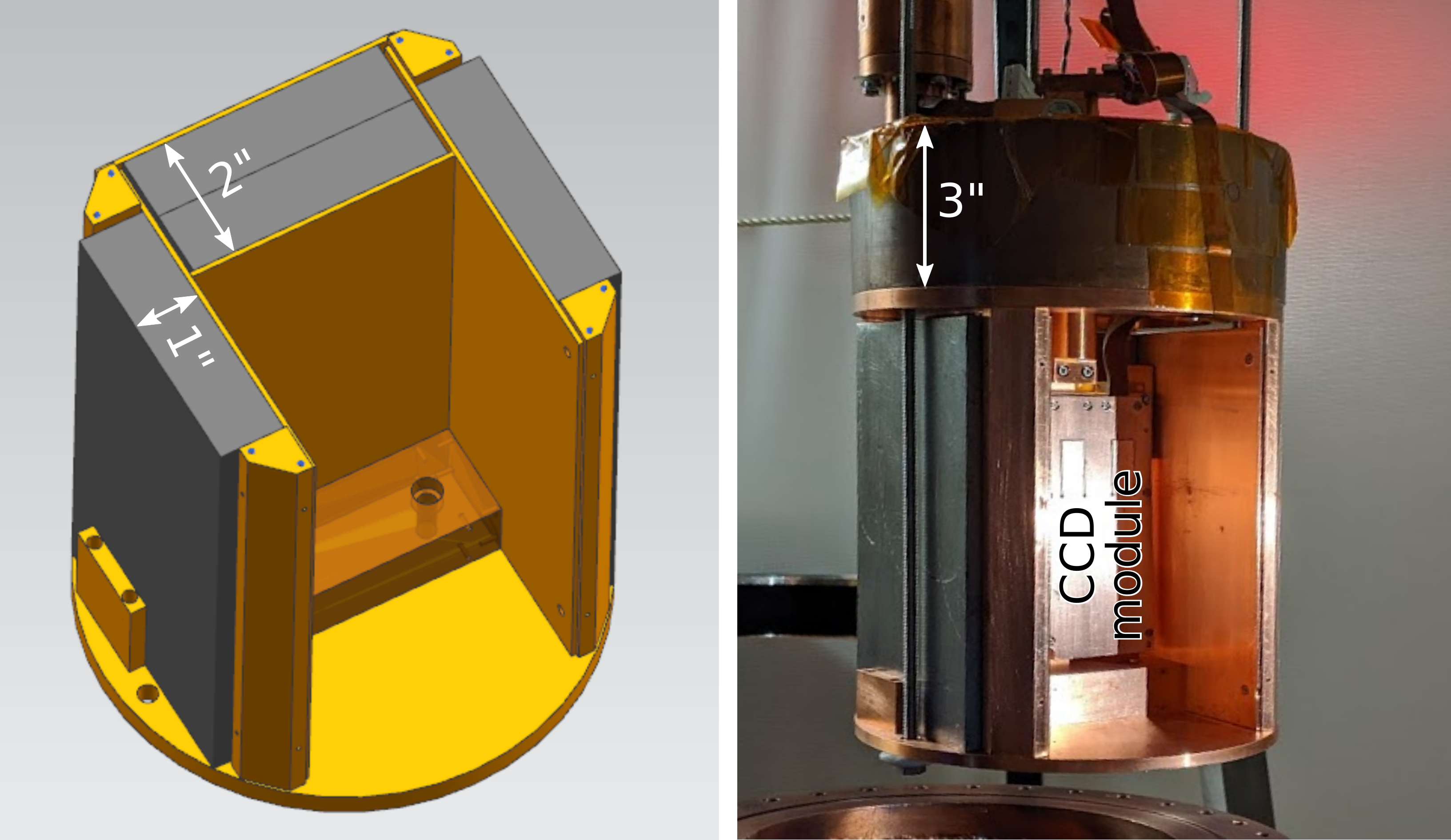}
\caption{Diagram of the {\it standard} shield with the 2-inch lead cover removed to show the inside (\textbf{left}). Picture of the open {\it standard} shield with the Skipper-CCD module, from Fig.~\ref{fig:module} in the main text, installed (\textbf{right}). A 3-inch lead block above the sensor shields it from the cryocooler cold head and active electronics on the top of the vessel.}
\label{fig:innerShield}
\end{center}
\end{figure}

\begin{figure}[t!]
\begin{center}
\includegraphics[width=0.79\textwidth]{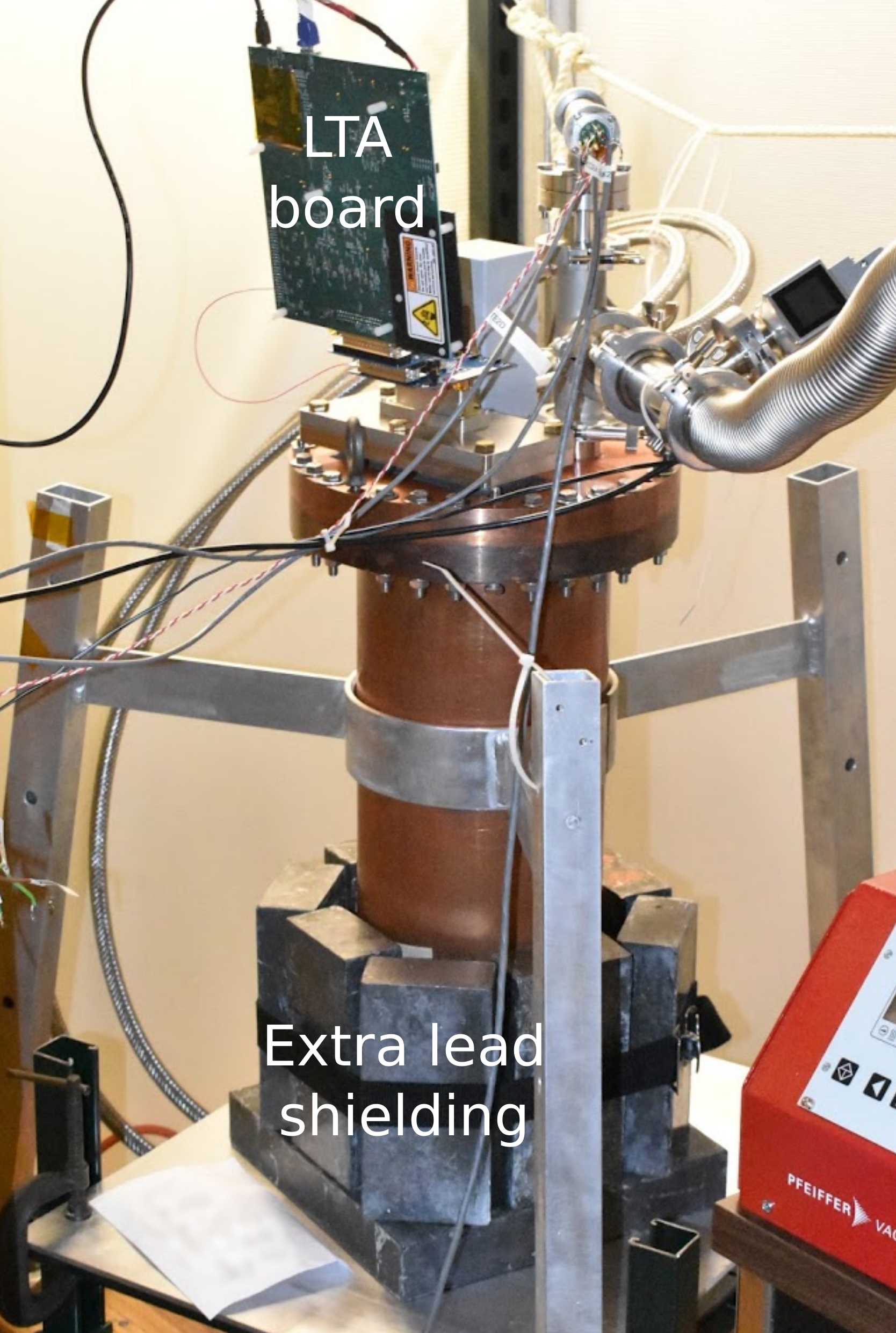}
\caption{The experimental apparatus with {\it extra} lead shielding installed. The top of the CCD module is located about 2~inch below the top surface of the exposed $2~{\rm in.} \times 4~{\rm in.} \times 8~{\rm in.}$ lead bricks. 
\vspace{-5mm}}
\label{fig:setup}
\end{center}
\end{figure}

\noindent\textbf{DETECTOR LAYOUT.}
A schematic drawing and a picture of the shield and vessel in which the Skipper-CCD module was placed is shown in Fig.~\ref{fig:innerShield}.  A photo of the entire experimental apparatus, showing also the extra lead shielding placed on the outside of the vessel and the LTA electronics board, is shown in Fig.~\ref{fig:setup}. 
\vskip 3mm

\noindent \textbf{SPURIOUS CHARGE.}
Several efforts have been made over the past months to understand the origin of the single-electron events. One of these efforts was focused on developing a method to measure the spurious charge (SC) generated during readout. Spurious charges are generated during the clocking of each pixel, either when they are being vertically moved in the bulk, or horizontally shifted across the serial register~\cite{janesick2001scientific,Haro:2016}. We note that the SC generated in the Skipper stage is negligible: such charges would appear partway through the readout of a charge packet and fill the space between the Gaussian peaks of the measured charge distribution. We do not observe this even in data taken with a large number of samples.

In contrast to true dark current, radiation-induced events, or DM signal events, the SC does not depend on exposure time.  
Instead, it depends on the number of times a pixel's surface is clocked. 
In December 2019, in the MINOS cavern, we measured the SC using the same Skipper-CCD as used for the DM search presented in this work, and at the same CCD temperature of 135~K.  

\begin{figure}[t!]
\begin{center}
\includegraphics[width=0.99\textwidth]{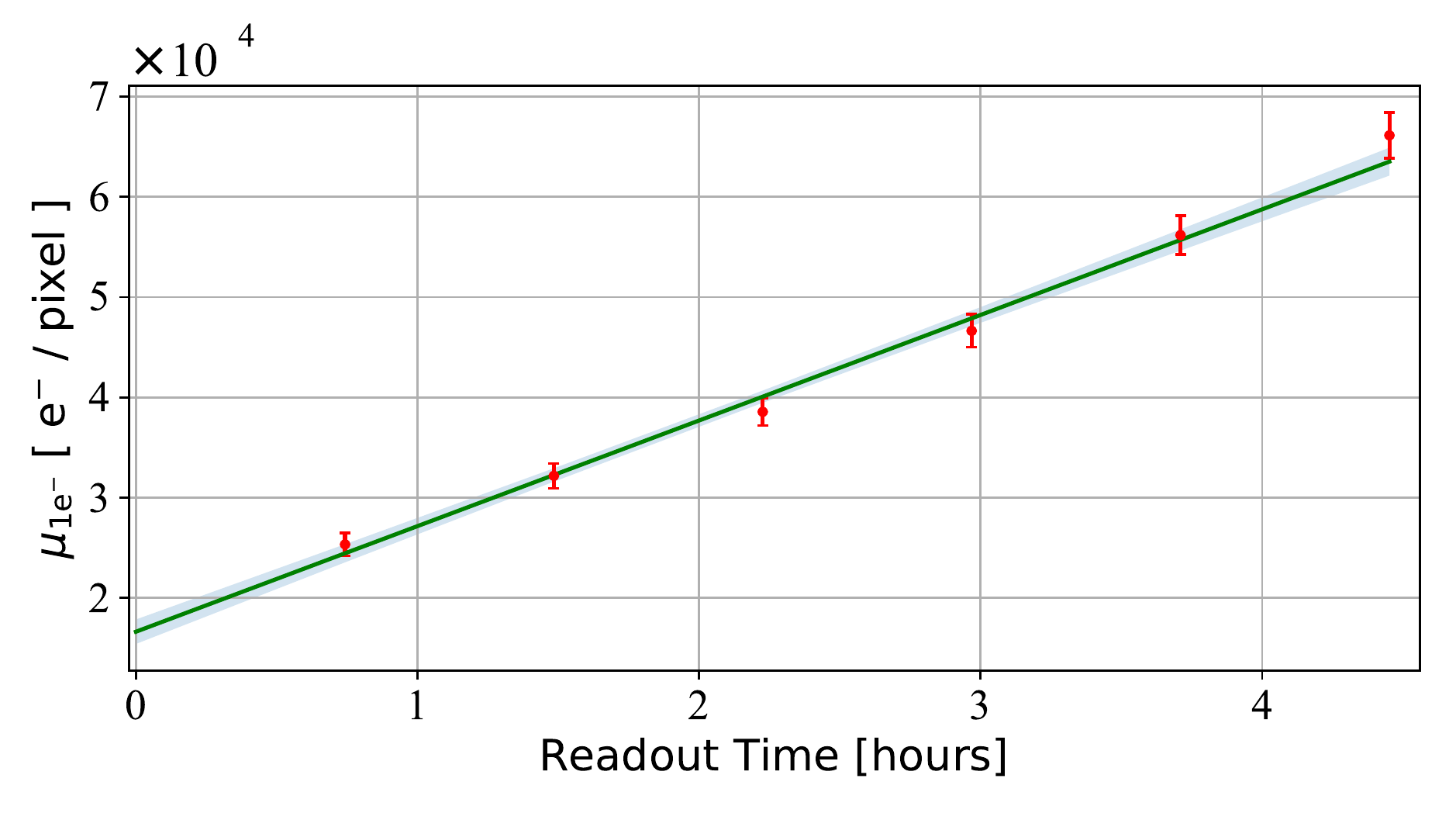}
\caption{Number of observed single-electron events (including those generated from spurious charge and other processes) expressed as $\mu_{1e^{-}}$ versus the readout time.  Since there is no exposure before readout, the effective exposure time is half the readout time and the $y$-intercept is an estimate of the SC.  The measured values with their statistical uncertainty are in red. A linear fit is in green, together with a $1\sigma$~CL band in light green.  
\vspace{-3mm}
}
\label{fig:SC}
\end{center}
\end{figure}

To measure the SC (for details see~\cite{Spurious-charge}), we expose the CCD for 0 seconds and then read it out at different speeds by varying the number of samples per pixel. The resulting readout time, $t$, ranges from $\sim$1~hour to $\sim$4.5~hours.  In addition to accumulating single electrons from the SC, the average pixel accumulates single electrons at a rate \Rs\ for a time $0.5t$. For each measurement, the number of observed single-electron events per pixel, $\mu_{1e^{-}}$, are extracted.  The value of $\mu_{1e^{-}}$ is related to \Rs\ and the SC as
\begin{equation}
    \mu_{1e^{-}} = 0.5 R_{1e^-} t + {\rm SC}\ .
\end{equation}
A linear function is then fit to this quantity versus the readout time, see Fig.~\ref{fig:SC}. The $y$-intercept of this linear regression, $(1.664 \pm 0.122)\times 10^{-4}$\e/pix, provides an estimate of the SC.

\begin{figure}[t!]
\begin{center}
\includegraphics[width=0.99\textwidth]{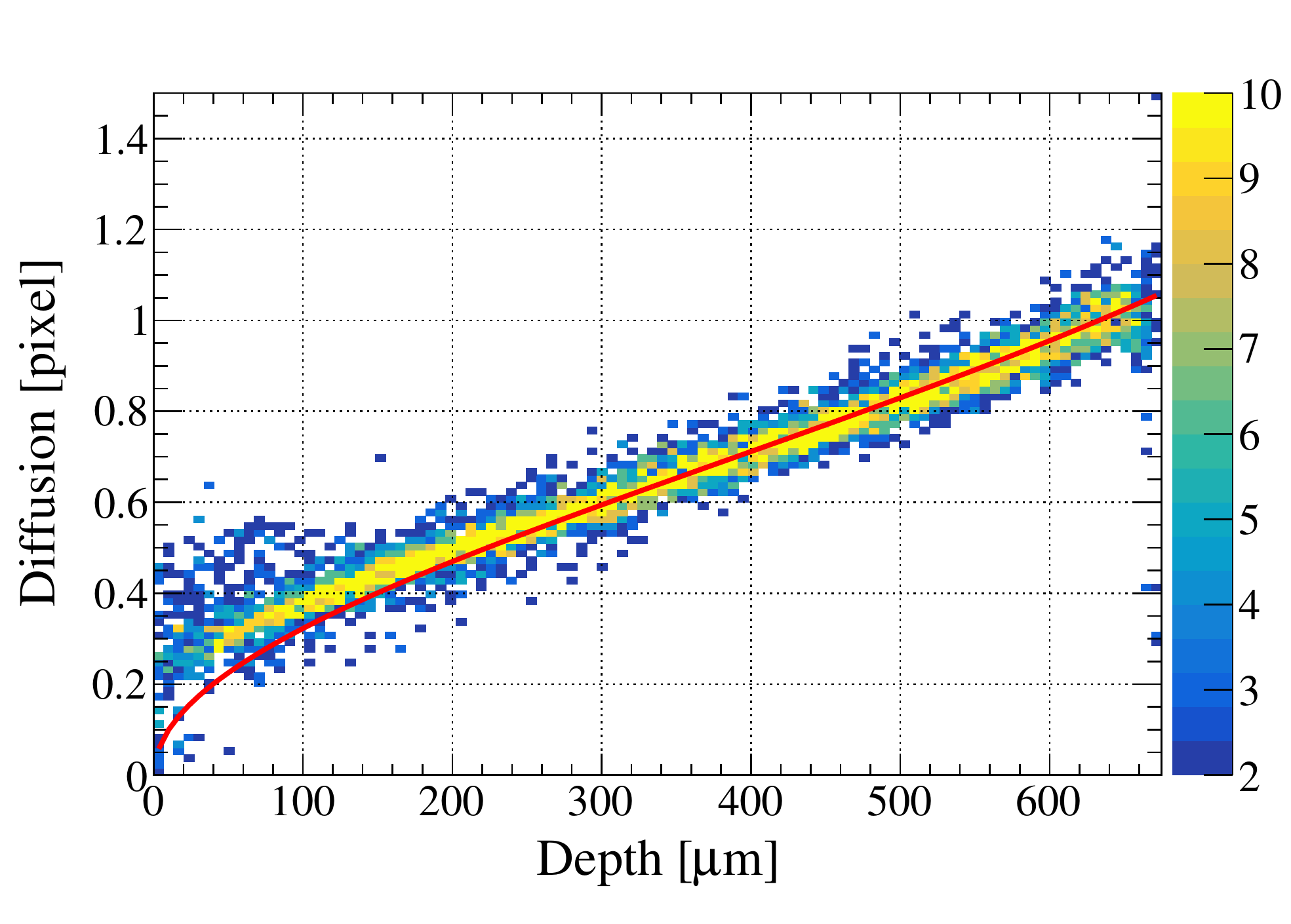}
\caption{We use 85 muons in the DM science data to measure the diffusion in the Skipper-CCDs.  Color scale shows the number of muons per bin in depth.  The solid line is the best fit from the diffusion model (see text for details).}
\label{fig:diffusion_calibration}
\end{center}
\end{figure}

\vskip 3mm
\noindent\textbf{DIFFUSION.}
The mapping of the diffusion of charge packets in the sensor as a function of the ionization depth was calibrated with atmospheric muons crossing the CCD sensor. An atmospheric muon produces a straight line signature with different widths given by the diffusion of the ionization from  different depths. This width-depth dependence can be measured and used to fit the expected diffusion model for thick CCDs~\cite{Holland:2003}. This is a standard strategy to calibrate the size of events in thick CCD sensors for particle detection~\cite{Aguilar-Arevalo:2016ndq, CONNIE_2019}. Fig.~\ref{fig:diffusion_calibration} shows the result using 85 muons extracted from the science data. The plot shows the measured diffusion as a function of depth with the color scale together with the best fit for depth points between 300~$\mu$m and 650~$\mu$m in the solid line. For shallower depths (below 300~$\mu$m), the small diffusion and the spatial quantization of the pixels masks the transport process of the charge packets, and the measurements deviate from the theoretical model. For depths above 650~$\mu$m some deviation from the theoretical model is seen due to the back side treatment of the sensor. Charge packet transport in thick sensors has been extensively studied in~\cite{Haro_2020}.

From the data in Fig.~\ref{fig:diffusion_calibration}, we find that the one-sigma diffusion versus depth is described by the function 
$\sigma = \sqrt{-A\log|1-bz|}$~pix, with $A=218.715$~$\mu$m$^2$ and $b=1.015\times 10^{-3}/\mu$m. 
We can use this to calculate the probability for $n$ electrons in a single pixel to diffuse to neighboring pixels and create a particular pattern. 
We only look at patterns where all electrons end up in connected pixels, or the same pixel, and calculate individual probabilities for these patterns. For this, we perform a Monte-Carlo simulation, in which we simulate the diffusion of $n$-electron events that originate at the same point within a pixel, and calculate the probability for each pattern. 
The resulting charge diffusion efficiencies relevant for the DM analyses presented in this paper for 1\e, 2\e, 3\e, and 4\e events are given in Table~\ref{tab:eff} in the main text. 

\begin{figure}[t!]
\begin{center}
\includegraphics[width=0.99\textwidth]{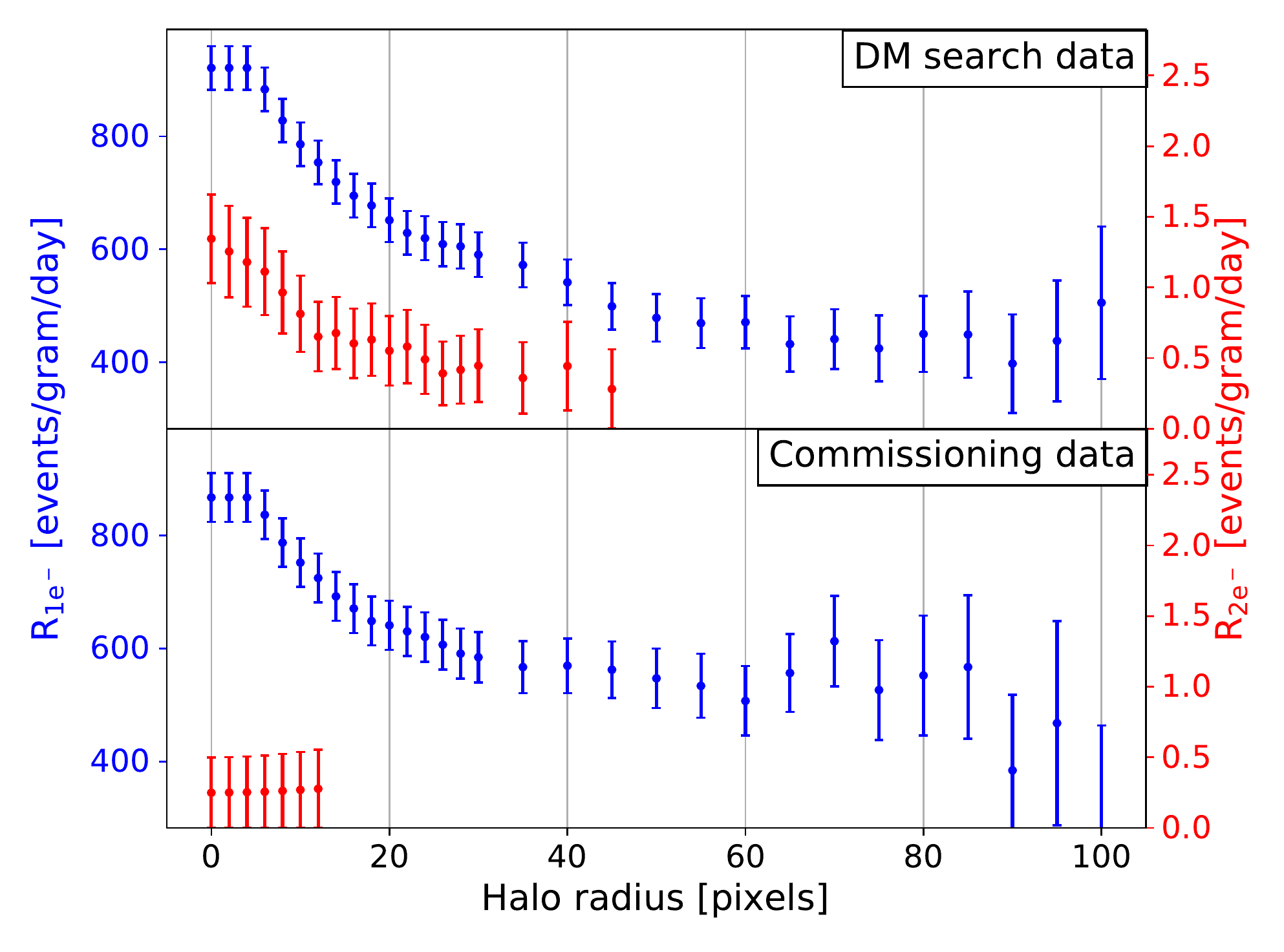}
\caption{Single-electron (\Rs, blue, left axis) and single-pixel two-electron (\Rtwo, red, right axis) event rates versus radius of the applied halo mask, centered on pixels containing more than 100\e (indicating a high-energy event) in commissioning (bottom) and DM search data (top). All event selection criteria discussed in the text are applied in this plot.}
\label{fig:Halo}
\end{center}
\end{figure}

%
\noindent\textbf{MASKING CUTS.}
Masking cuts were determined using (7) 20-hour-exposure commissioning images. These 7 commissioning images consist of 3 images that have an identical readout scheme as the DM science data, and 4 images in which the entire CCD was read through the amplifiers in quadrant-1 and quadrant-2 only. 
During readout, the first 8 columns in each row are part of a non-active ``prescan'' region; we also read a 19-column-wide ``overscan'' region to check the CCD noise. 
We provide a few additional details and explanations here that were not in the main text. 
\begin{itemize}[leftmargin=*]\addtolength{\itemsep}{-0.6\baselineskip}
\item {\bf Charge Diffusion.} While we consider DM events that occur in multi-contiguous-pixel events for the 3\e and 4\e analyses, we consider only single-pixel DM events for the 2\e analysis, since \Rs\ is sufficiently high to create random coincidences between neighboring pixels.  
\item {\bf Readout Noise.} 
We veto images in which the readout noise is 30\% larger than the expected readout noise as inferred from an over-scan region in which virtual (non-existent) pixels are read. 
\item {\bf Crosstalk.} 
High-energy signals recorded in one of the four quadrants can produce a fake signal in one or more of the other three due to ``crosstalk''~\cite{Bernstein:2017gsy}. Any pixel is masked if it is read at the same time as a pixel containing more than 700~electrons (which produces about $-0.1$\e of crosstalk).
\item {\bf Serial Register Hit.}
Serial register hits show up as a set of non-empty consecutive pixels in a particular row. We remove any row if it has five consecutive pixels with at least four pixels having $\ge$1\e, and the five pixels in the rows before and after each average $<$0.5\e. 
\item {\bf Low-Energy Cluster.} 
This cut masks regions where any additional $\le$4\e-events would be viewed as likely originating from backgrounds.  The 3\e and 4\e cuts are stricter (20~pix) than the 1\e and 2\e cuts (4~pix). The 3\e and 4\e cuts were developed after unblinding the 1\e and 2\e analyses, at which point we noticed regions containing multiple nearby (but disconnected) 2\e events.  These regions are typically removed also by the Loose Cluster and Nearest Neighbor cut, but not necessarily so, and so we implemented a stricter cut for the 3\e and 4\e analyses before unblinding these data.  
\item {\bf Edge Mask.}  We remove 60 (20) pixels around all edges of a quadrant for the 1\e ($\ge$2\e) analyses. 
\item {\bf Bleeding Zone Mask.}
We mask 100 (50) pixels upstream in the vertical and horizontal direction of any pixel containing more than 100\e for the 1\e ($\ge$2\e) analyses. 
In addition, we define a ``bleed rate'' for a particular column as the number of electrons in the bleeding-zone pixels (after applying an Edge Mask of 20 pixels, the Crosstalk, Serial Register Hit, and Low-Energy Cluster Halo masks) of that column divided by the number of such pixels, where the numbers of electrons and pixels are summed across all images. 
We double the number of pixels masked along a column (to 200 (100) pixels for the 1\e ($\ge$2\e) analyses) in those columns where the bleed rate is more than 3.71 median absolute deviations above the median bleed rate across all columns of the current quadrant. 
\item {\bf Bad Pixels and Bad Columns.} 
We remove several pixels that were observed to have an unusually large \Rs\ (orders of magnitude larger than in typical pixels) when taking data at 210~K, indicating the presence of defects or impurities that cause charge leakage (so-called ``dark spikes'')~\cite{janesick2001scientific}. 
In the data analysis, we apply an Edge Mask of 20 pixels and the Crosstalk, Serial Register Hit, and Low-Energy Cluster Halo masks to the DM search data, and identify pixels and columns that have a significant excess of charge.
In particular, we remove any pixels that have at least 1\e three times or have at least two events with a total of at least 3\e (e.g., one 1\e and one 2\e event, or two 2\e events). 
Similarly, after applying the above masks and the Bleeding Zone Mask, we count the number of electrons in all pixels in a column summed over all images and divide by the number of such pixels, and remove any columns for which this rate is more than twice the rate averaged over all columns of the current quadrant, and any columns that contain at least two 2\e events.
Based on the measured \Rs, we would expect this mask to remove 0.16 normal pixels and 0.003 normal columns due solely to statistical fluctuations.
\item {\bf Halo Mask.}  We remove 60 (20) pixels around any pixel with more than 100\e for the 1\e ($\ge$2\e) analyses.  The 1\e analysis is not exposure-limited, so a stringent cut of 60 pixels is acceptable.  For the 2\e analysis, removing 20 radial pixels is sufficient to remove nearby 2\e events while still maintaining a high signal efficiency for obtaining a near-optimal 2\e constraint.  We keep the same 20 radial pixel cut for the 3\e and 4\e analyses.  
Fig.~\ref{fig:Halo} shows the radial decrease in \Rs\ in the commissioning and DM search data.  
The value of \Rs, in blue, shows a monotonic decrease out to $\sim$35 pixels, while \Rtwo, in red, remains constant (corresponding to 1 observed event) until 0 events are counted for a radius of 12 pixels. 
\item {\bf Loose Cluster Mask.} 
Upon further analyzing the data we had collected with the prototype detector~\cite{Abramoff:2019dfb}, as well as in the commissioning data for this work,  we see a correlation between the number of 1\e and 2\e events in small regions on the Skipper-CCD, of size $\sim\ 10^3$~pix$^2$.  
These regions could arise from, e.g., showers of nearby background events outside of the active area of the Skipper-CCD. 
After applying the previous masks, we mask regions that contain an excess of 1\e events as follows: for each 1\e pixel, if there is another 1\e pixel within a radius of 20~pix, we remove all pixels in a 20-pixel radius. We apply this mask only for the $\ge$2\e analyses. For the 3\e and 4\e cluster analyses, only 1\e events passing the Neighbor Mask (described below) are used to apply this mask. 
\item{\bf Neighbor Mask.} For the 1\e and 2\e DM analyses only, we require the DM signal to be contained in a single pixel and only select pixels whose eight neighboring pixels are empty. We thus mask all pixels that have a neighboring pixel with $\ge1\e$.
\item{\bf Geometric Efficiency for Clusters.} For the 3\e and 4\e DM analyses where we accept clusters of multiple contiguous pixels, the width of the area where a DM event could be detected is reduced by the width of the cluster. We apply this correction for every possible cluster shape, and incorporate it in the efficiency-corrected exposure.
\end{itemize}%

Note that we recompute the total exposure times efficiency after applying all event-selection criteria; since each pixel has a unique exposure, the final efficiency-corrected exposure is not simply the exposure before masking times the total cut efficiency.

\vskip 3mm
\noindent\textbf{SINGLE-ELECTRON EVENT RATE VERSUS CCD TEMPERATURE.}
Thermal generation of electron-hole pairs, mediated by defects or impurities, is a well-understood source contributing to the single-electron rate, \Rs, in CCDs. This is typically referred to as ``dark current'' (here abbreviated as DC) and has a temperature dependence that is well understood~\cite{janesick2001scientific}. In fully depleted CCDs such as the SENSEI Skipper-CCDs, two types of DC are expected to dominate: bulk DC that originates from defects in the depleted bulk silicon, and surface DC that originates from defects at the silicon-oxide interface. Both typically scale with the same temperature dependence. 
The surface DC can be temporarily suppressed by an ``erase'' procedure that brings the surface into inversion, but recovers to its equilibrium value with a time constant that increases at lower temperature~\cite{69907, Holland:2003}.

\begin{figure}[b!]
\begin{center}
\includegraphics[width=0.99\textwidth]{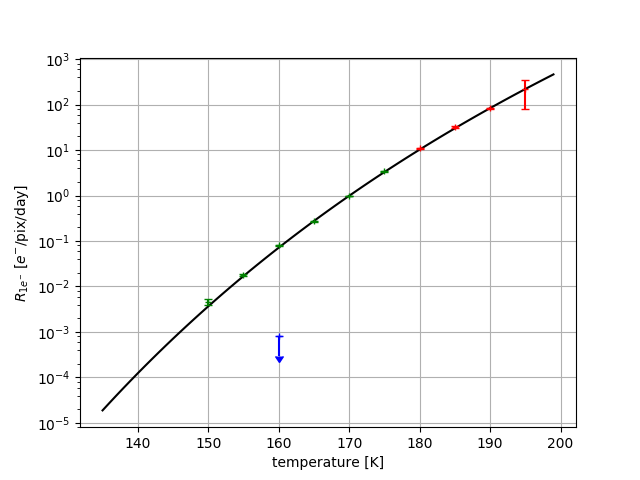}
\caption{Dependence of \Rs on temperature. The measurements from 150 to 175~K in blue (green) are from a CCD at SNOLAB, and taken with (without) erasing the surface dark current. The measurements from 180 to 195~K (red) are from a CCD at the surface, and taken without erasing the surface dark current. The black curve is a fit to the green and red points, using the theoretical model for the temperature dependence of dark current~\cite{janesick2001scientific}.
\vspace{-5mm}
}
\label{fig:dc_vs_t}
\end{center}
\end{figure}

We measured \Rs\ in two science-grade Skipper-CCDs from the same production batch as the one used for the DM search. One was operated at the surface, and the other was installed at the SNOLAB deep underground facility.

As shown in Fig.~\ref{fig:dc_vs_t}, measurements of \Rs\ without erase (red and green points) follow the expected temperature dependence (black fit curve) down to 150~K.
This should represent the sum of bulk DC and equilibrium surface DC.
At 160~K we put an upper bound (blue point) on \Rs\ of $8 \times 10^{-4}$~\e/pix/day immediately after erase, as compared to $(8.0\pm 0.5)\times 10^{-2}$~\e/pix/day without erase. The bulk DC is therefore at least a factor of 100 smaller than the surface DC, consistent with our expectation given the high quality of the silicon used.

Our best estimate for the DC contribution to \Rs\ is to extrapolate the black curve of Fig.~\ref{fig:dc_vs_t} to 135~K (thus accounting for the temperature dependence), and shift it down by the distance between the blue and green points at 160~K (thus accounting for the effect of the erase procedure). This yields an estimate of $<1\times 10^{-6}$~\e/pix/day, but with considerable uncertainties. Because this value is negligible relative to the measured \Rs, we conservatively choose not to subtract it from our final limit on \Rs.

\vskip 3mm
\noindent \textbf{DATA WITH STANDARD SHIELD.}
The data with the standard shield, for which we see a higher \Rs\ as described in the main text, consist of four images, each of which has a 12-hour exposure.  The number of samples per pixel is 400.  One image has the output transistor of the amplifier turned off during exposure, while the other three images have the output transistor of the amplifier turned on. The data were otherwise taken with the same CCD settings as the DM science data. Since we find no evidence in the new high-resistivity Skipper-CCDs for amplifier-induced events that are present for our prototype detectors~\cite{Abramoff:2019dfb}, we expect the amplifier-on images to provide a reliable measurement of \Rs.  In any case, the amplifier-on images have a single-electron event rate of $R_{1e^-}=(4.302^{+1.743}_{-1.426})\times 10^{-4}$~\e/pix/day, which is smaller than the value for the amplifier-off image, namely $R_{1e^-}=(7.555^{+3.286}_{-2.562})\times 10^{-4}$~\e/pix/day. 
Averaged over the four images, we find $R_{1e^-}=(5.312^{+1.490}_{-1.277})\times 10^{-4}$~\e/pix/day for the standard shield. 

The measured value of \Rs\ in the standard-shield data is to be compared with the
value measured in the DM science data that have the extra lead, $R_{1e^-}=(1.594 \pm 0.160)\times 10^{-4}$~\e/pix/day. 
Assuming, conservatively, a $3\sigma$ upward fluctuation in the estimated SC contribution to both the DM science data and the combined standard-shield data, we find that the probability~\cite{KRISHNAMOORTHY200423} of obtaining the measured values of the two data sets under the assumption that \Rs\ is equal under both conditions is $4.70\times 10^{-4}$ ($3.3~\sigma$). (Assuming an upward fluctuation in the SC contribution makes the probability estimate conservative, since the standard-shield data have a shorter exposure time than the DM science data and therefore, comparatively, a larger contribution to the number of 1\e events from SC.) 
In the future, in order to verify and characterize the correlation between the high-energy event rate and \Rs, we plan to take more standard-shield data, data with other shield thicknesses, and data with radioactive sources (such as Co-60) at different distances from the CCD.

\begin{figure}[t!]
\begin{center}
\includegraphics[width=0.99\textwidth]{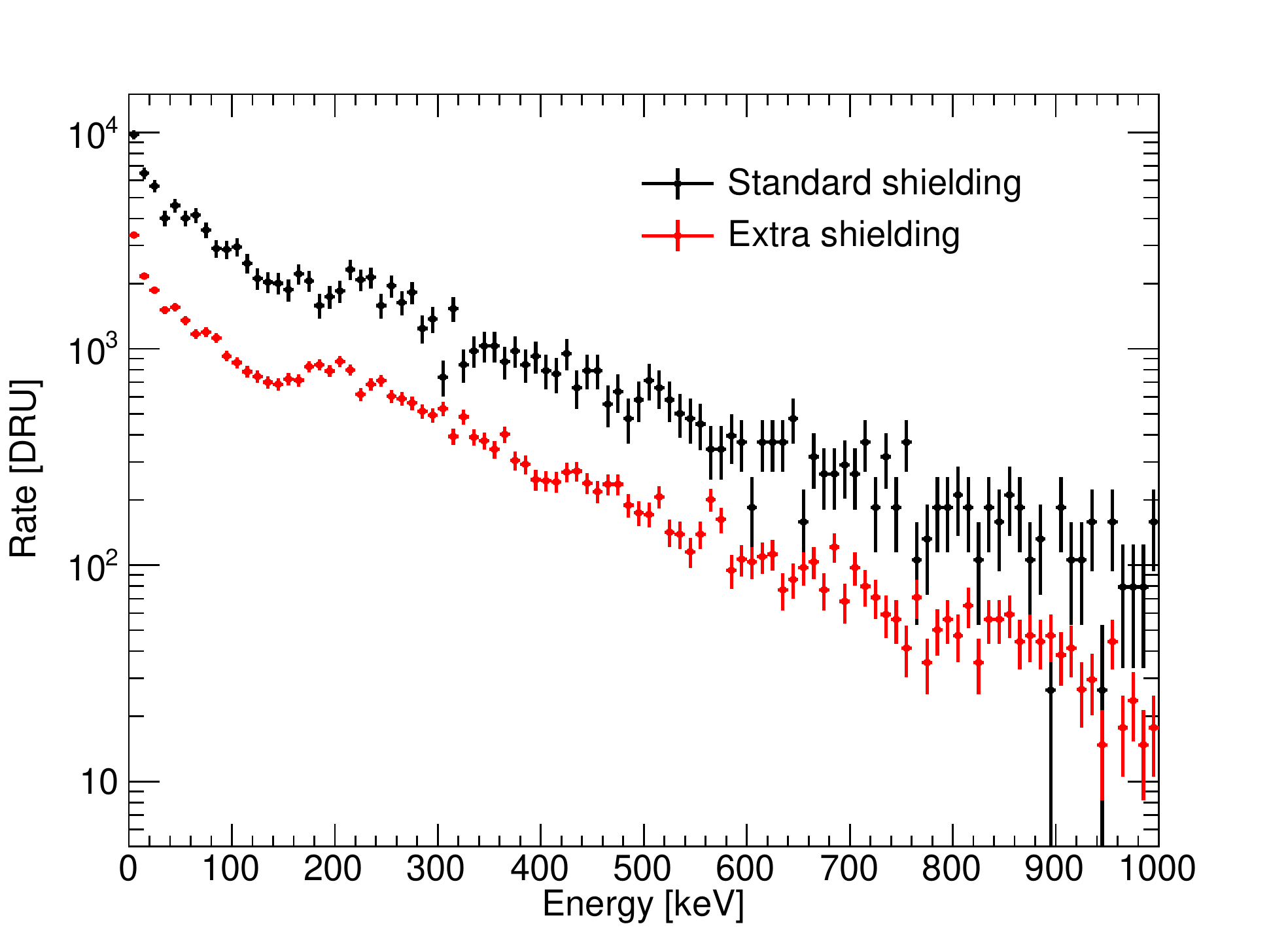}
\caption{High-energy event spectra from 10~keV to 1~MeV for data taken with the standard shield (black) and for data taken with the extra lead shield (red). 
\vspace{-5mm}
}
\label{fig:HEspectra}
\end{center}
\end{figure}

\vskip 3mm
\noindent \textbf{HIGH-ENERGY SPECTRA.}
We show the high-energy event spectra from 10~keV to 1~MeV in Fig.~\ref{fig:HEspectra}, for data taken with the standard shield and for data taken with the extra lead shield.

\begin{figure}[b!]
\begin{center}
\includegraphics[width=0.49\textwidth]{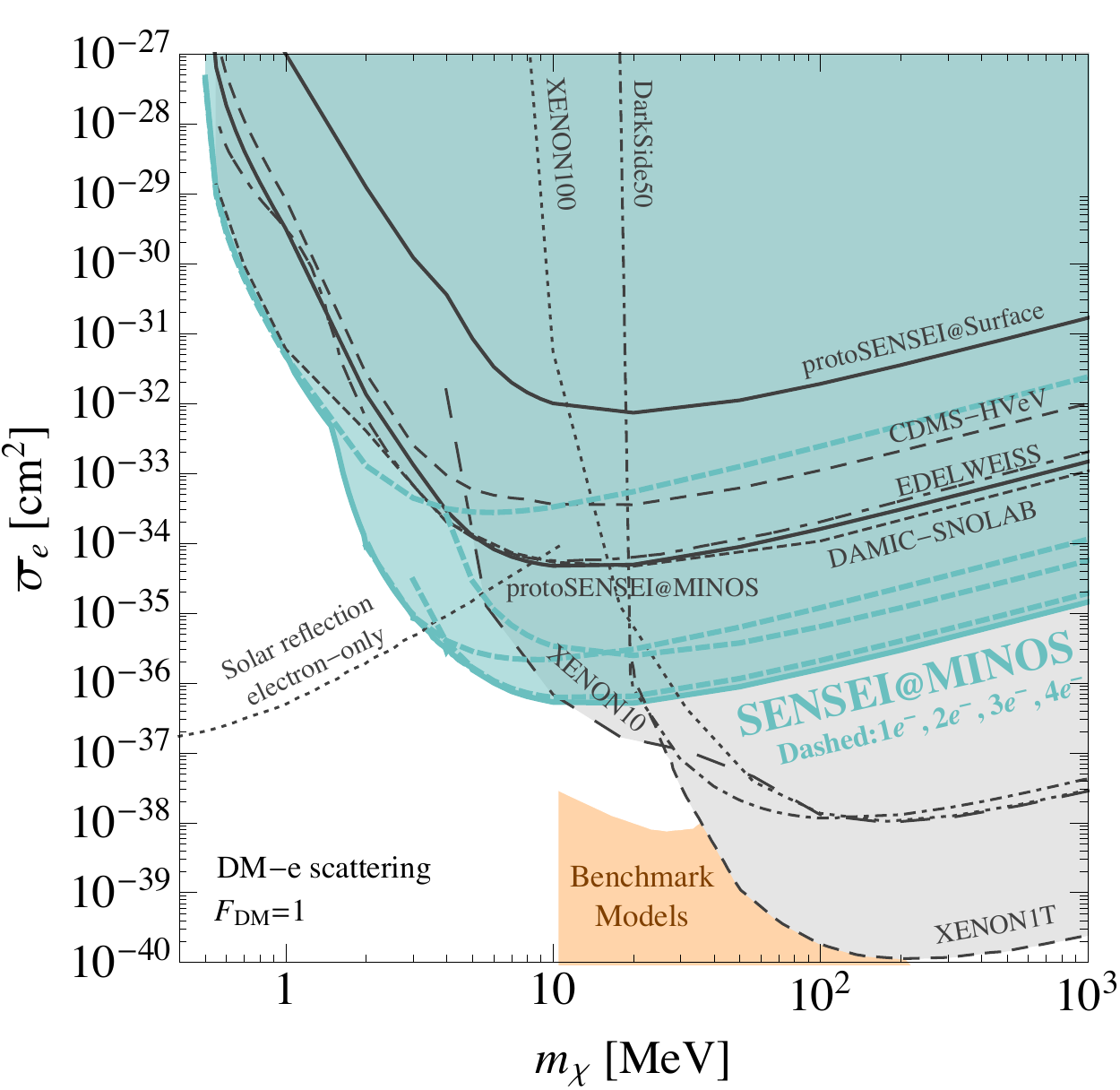} 
 \includegraphics[width=0.49\textwidth]{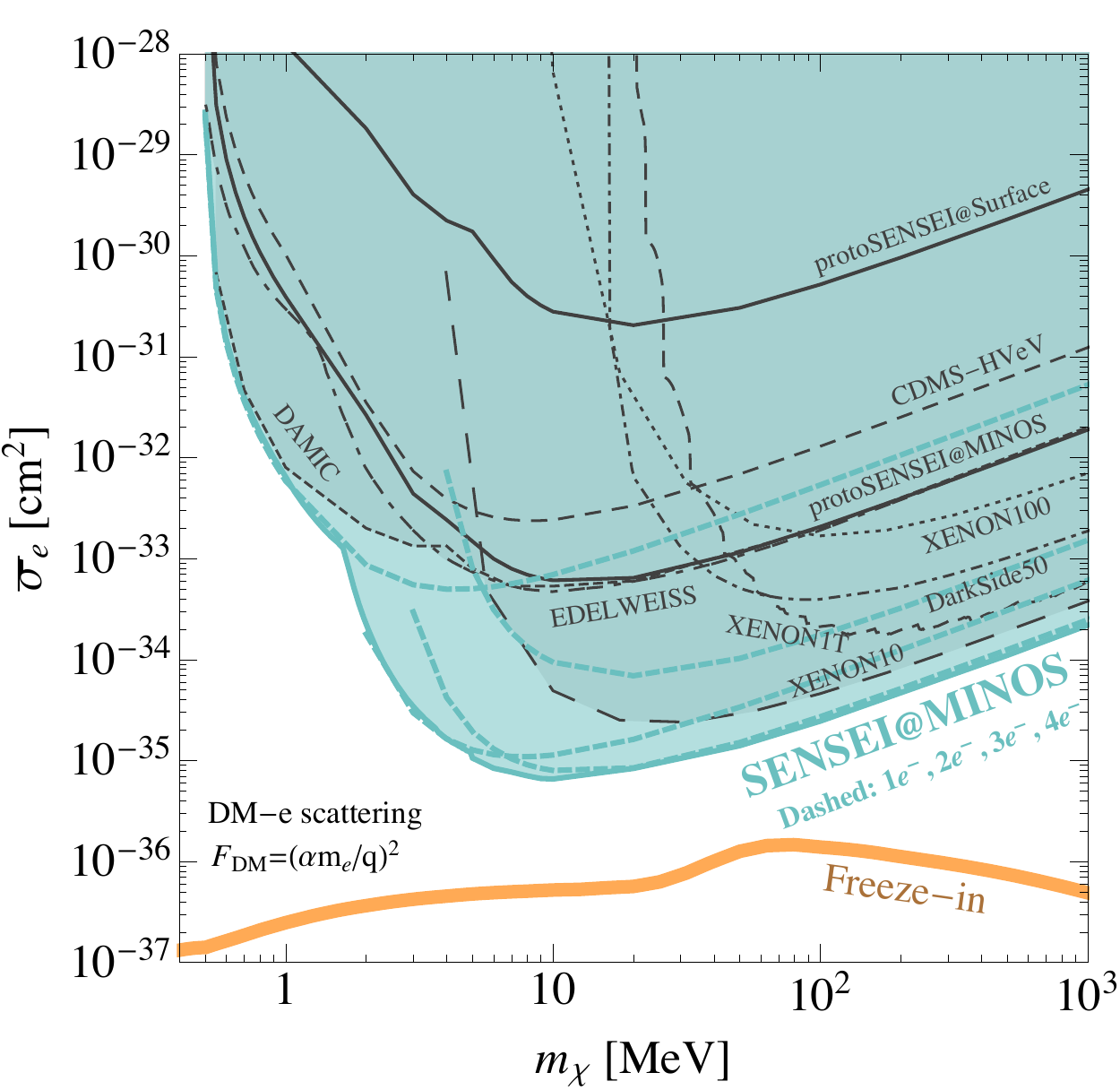}
    \caption{
90\%~CL~constraints on: DM-\e cross section, $\overline{\sigma}_e$, 
versus DM mass, $m_\chi$, for two DM form factors, $F_{\rm DM}(q)=1$ (\textbf{left}) and $F_{\rm DM}(q)=(\alpha m_e/q)^2$ (\textbf{right}). 
Cyan dashed lines correspond to limits from \textit{individual} $N_e =$~1\e, 2\e, 3\e, and 4\e  analyses (shifting to higher masses for higher $N_e$), while the solid line corresponds to the 90\% c.l.~limit calculated using a likelihood-ratio test that combines all four bins.  Other constraints are identical to those presented in the letter. 
 }
\label{fig:DMresultsIndividualLines}
\end{center}
\end{figure}

\newpage 
\noindent \textbf{COMPARISON OF LIMIT IN LETTER WITH LIMIT FROM INDIVIDUAL ELECTRON BINS.}
In Fig.~\ref{fig:DMresultsIndividualLines}, we show the 90\% c.l.~DM limits calculated from each bin individually, and compare this with the 90\% c.l.~DM limit calculated with a likelihood-ratio test based on~\cite{Cowan:2010js} and presented in the main part of the letter.  We see that the likelihood-ratio test limits are very slightly stronger than the limits calculated from the individual bins. We note that the XENON10 limit is calculated using individual bins as described in~\cite{Essig:2017kqs}. 

\end{document}